\documentclass[10pt, final]{IEEEtran}

\usepackage{amsthm}

\usepackage{amssymb}
\usepackage{amsmath}
\usepackage{bbm}
\usepackage{mathrsfs}
\usepackage{cite}
\usepackage{bm,epsfig,amsthm,url}
\usepackage{indentfirst}
\usepackage{amsfonts}
\usepackage{epstopdf}
\usepackage{cases}
\usepackage[caption=false,font=scriptsize]{subfig}
\usepackage{xcolor}
\usepackage{mathtools}

\makeatletter
\renewcommand{\maketag@@@}[1]{\hbox{\m@th\normalsize\normalfont#1}}%
\makeatother
\usepackage{hyperref}
\usepackage{stfloats}

\newtheoremstyle{mystyle}{}{}{}{}{}{: }{0pt}{\indent \it{\thmname{#1}\thmnumber{ #2}\thmnote{#3}}}
\theoremstyle{mystyle}

\newtheorem{Proposition}{Proposition}

\usepackage[noend]{algpseudocode}
\usepackage{float}
\usepackage{algorithmicx}
\usepackage[linesnumbered,ruled,vlined]{algorithm2e}

\allowdisplaybreaks[4]

\begin{document}
	
\title{\fontsize{22pt}{26pt}\selectfont Intelligent Reflecting Surface for ISAC: From Beamforming to Joint Design With Surface Orientation}
	
\author{{Qiaoyan~Peng, Qingqing~Wu, Guangji~Chen, Yong~Zeng, Sai~Huang, Zhiyong~Feng, Shanpu~Shen, Wen~Chen}
\thanks{Q. Peng is with Foshan University, Foshan 528225, China, and also with the School of Integrated Circuit, Shanghai Jiao Tong University, Shanghai 200240, China (email: pengqiaoyan@fosu.edu.cn). 
	   Q. Wu and W. Chen are with the School of Integrated Circuit, Shanghai Jiao Tong University, Shanghai 200240, China (email: qingqingwu@sjtu.edu.cn; wenchen@sjtu.edu.cn).
	   G. Chen is with Nanjing University of Science and Technology, Nanjing 210094, China (email: guangjichen@njust.edu.cn).
       Y. Zeng is with the National Mobile Communications Research Laboratory, Southeast University, Nanjing 210096, China, and also with Purple Mountain Laboratories, Nanjing 211111, China (e-mail: yongzeng@seu.edu.cn)
       S. Huang and Z. Feng are with the Key Laboratory of Universal Wireless Communications, Ministry of Education, Beijing University of Posts and Telecommunications, Beijing 100876, China (e-mail: huangsai@bupt.edu.cn; fengzy@bupt.edu.cn).
	   S. Shen is with the State Key Laboratory of Internet of Things for Smart City, University of Macau, Macao 999078, China (email: shanpushen@um.edu.mo).}
}
	
\maketitle
	
\begin{abstract}
Rotatable semi-passive intelligent reflecting surfaces (SIRSs) provide an additional spatial degree of freedom (DoF) by proactively adjusting the orientation, while dedicated sensors directly receive target echoes. To exploit this DoF for integrated sensing and communication (ISAC), we investigate the joint optimization of SIRS rotation, base station (BS) transmit beamforming, and SIRS phase-shifts to minimize the Cramér–Rao bound (CRB) for target's spatial-frequency estimation subject to per-user communication signal-to-interference-plus-noise ratio (SINR) constraints. We first establish an orientation-dependent system model that characterizes the effective aperture gains, based on which the CRB is derived. To gain useful insights, closed-form expressions of the beamforming and rotation designs are derived in the sensing-only and communication-only cases. For the co-located user and target case, we further characterize the condition under which the sensing-oriented rotation design also satisfies the communication requirement. In the single-user single-target case with ideal unit aperture gains, we analyze the channel correlation and its impact on ISAC performance. For the general multi-user case, an effective method based on the genetic algorithm is developed. Numerical results demonstrate the superiorities of the rotatable SIRS over the fixed-orientation SIRS and show that jointly optimizing the SIRS orientation and beamforming can improve the effective aperture gains and the correlation between communication and sensing channels.
\end{abstract}
	
\section{Introduction}
Integrated sensing and communication (ISAC) has been recognized as a key usage scenario for sixth-generation (6G) wireless networks \cite{luo2026ISACSurvey}. By sharing spectrum, hardware, and transmit waveforms, ISAC supports communication and sensing on an integrated platform, enabling applications such as autonomous driving, intelligent transportation, and industrial automation \cite{huangUAV}. However, communication and sensing generally impose distinct requirements on transmitted signals. Specifically, communication requires reliable signal reception at users and the suppression of inter-user interference, whereas sensing requires sufficient target illumination power \cite{niu2025InterferenceSurvey}. The resulting trade-off is governed by the correlation between communication and sensing channels, which motivates the use of additional spatial degrees of freedom (DoFs) \cite{bian2025Correlation,hua2024MIMOTradeoff,ren2024MultitargetTradeoff,feng2026ISAC}.

The performance of ISAC systems is critically dependent on the wireless propagation environment. In the absence of direct line-of-sight (LoS) links, both the received communication signal power and target illumination may be severely degraded, making it difficult to ensure reliable communication and ubiquitous sensing through base station (BS) transceiver design alone. To address these challenges, intelligent reflecting surface (IRS) has emerged as a cost-effective technology for reconfiguring wireless propagation environments by adjusting their reflection coefficients \cite{tishchenko2025HybridSurvey,rihan2024PassiveActive,ma2024RISISAC,ji_RDARS_model_driven}. By creating additional reflected paths and providing passive beamforming gains, IRSs can enhance the received communication power and target echo signals. 

Motivated by these benefits, IRS-aided ISAC has been investigated in prior studies. Depending on their hardware configurations, IRS architectures for ISAC can be categorized into passive IRS, active IRS, and semi-passive IRS (SIRS). For passive IRS-assisted ISAC with sensing echo received at the BS, the sensing signal propagates along the BS-IRS-target-IRS-BS path, thereby suffering severe cascaded propagation loss. A multi-hop IRS-enabled ISAC system was studied in \cite{luo2026MultihopPositioning}, where transmit beamforming and IRS phase-shifts were jointly designed to enhance target positioning accuracy while satisfying user communication constraints. For single-IRS and dual-IRS architectures, \cite{sankar2024RISBeamforming} investigated joint beamforming and reflection design for radar beampattern matching and worst-case target illumination power maximization, respectively, subject to user signal-to-interference-plus-noise ratio (SINR) constraints. Furthermore, in \cite{liu2024SNRCRB}, the BS beamforming and IRS reflection were jointly optimized to maximize the communication sum-rate under sensing signal-to-noise ratio (SNR) or Cramér–Rao bound (CRB) constraints. To compensate for path loss, active IRS is equipped with active reflecting elements to amplify incident signals \cite{peng2026ActiveULDL}. Thanks to its reflection amplification capability, it has been exploited for local sensing \cite{han2025MultifunctionalRIS}, secure transmission \cite{hua2024SecureIRSISAC}, and ISAC \cite{zhu2024ActiveCRB,yigit2025HybridSTARISAC}. However,  it may also amplify undesired noise and interference. Another type of active IRS, also known as a self-sensing IRS, integrates passive reflecting elements with active sensors and controllers to support signal transmission, reception, and processing. A self-sensing IRS-aided mmWave ISAC system was investigated in \cite{selfSensingIRS}, where a two-phase transmission scheme was developed for joint location sensing and channel estimation, and IRS beamforming was optimized for CRB minimization. Despite these benefits, both types of active IRSs introduce additional power consumption and hardware complexity due to the active components.

Considering both the path loss and hardware cost, a SIRS is equipped with dedicated sensors that receive and process the target echo signal for parameter estimation via the BS-IRS-target-IRS link. In the literature, SIRS-enabled sensing has been studied through CRB analysis and joint beamforming design \cite{peng2024SemiPassive,chen2026ISAC}, joint time-of-arrival and direction-of-arrival estimation \cite{hua2024SemiPassiveLocalization}, as well as integrated user localization and communication \cite{hu2023LocationISAC}. In addition, for SIRS-assisted ISAC, energy efficiency maximization with communication and sensing SNR constraints was studied in \cite{rojith2026SemiPassiveEE}. The authors of \cite{dong2026AngleVelocity} considered joint angle and radial velocity estimation. Furthermore, CRB optimization with and without prior angle information were investigated in \cite{dong2025SemiPassiveNearField}. In these studies, beamforming and reflection coefficients are optimized with pre-defined IRS orientations. However, a small effective aperture caused by undesired incidence or reflection angles cannot be fully compensated for by phase alignment alone. This limitation motivates exploiting the IRS orientation as an additional spatial DoF.

To address the deployment issue and motivated by the rotatable antenna \cite{qu2025RotatableCooperativeISAC,wang2026RotatableHybridISAC,zhang2026RotatableNearField,sun2025RotatableMovableISAC,luRA}, joint IRS position and rotation optimization has been studied under angle-dependent channel models \cite{rotation_IRS,feng2025SixDIRS,peng2026CooperativeRotatableIRS,liu2025RotatableIRS,wang2025rotate}. Specifically, an IRS at a fixed location can adjust its incidence and reflection geometry through mechanical rotation. In \cite{peng2026CooperativeRotatableIRS}, cooperative IRS beamforming and orientation were jointly designed, with closed-form expression of solutions derived under line-of-sight (LoS) channels. Furthermore, rotatable IRSs have been experimentally evaluated for max-min area coverage in \cite{liu2025RotatableIRS}. The work \cite{wang2025rotate} investigated an unmanned aerial vehicle (UAV)-mounted passive IRS-enabled ISAC system, where the IRS location, orientation, and reflection coefficients were optimized jointly with the BS transmit beamforming to maximize the sensing SNR subject to a communication SNR requirement. These results highlight the importance of jointly designing IRS beamforming and orientation to improve the effective reception and reflection gains, and demonstrate the benefits of geometric reconfiguration beyond conventional phase-shift design. Nevertheless, the work \cite{wang2025rotate} considered a single-user ISAC system with the objective of sensing SNR, where the target echo signals are received at the BS through a fixed antenna array. In contrast, a SIRS directly collects the target echo signal through its sensors, whose orientation changes with the reflecting panel. As such, SIRS rotation affects not only target illumination but also the receive array response, introducing an additional design dimension for target parameter estimation. Moreover, in a multi-user ISAC system, the SIRS needs to assist the BS to serve users located in different directions while mitigating inter-user interference.

Regarding the deployment of SIRS in an ISAC system, several fundamental issues remain unaddressed. First, how to design a rotation scheme to flexibly balance between communication and sensing performance? SIRS rotation changes the effective aperture gains of the BS-IRS-user and BS-IRS-target links. An orientation that improves target performance may reduce the user signal strength, requiring the BS and SIRS beamforming to be adjusted to satisfy the communication requirement. Conversely, adjusting the SIRS rotation to strengthen the communication link may reduce the target illumination gain and degrade sensing performance. Second, how to characterize the impact of channel alignment on CRB for target's spatial-frequency estimation? Note that the CRB depends not only on the received echo power but also on the derivative of the target response with respect to (w.r.t.) the estimated parameter. SIRS rotation changes the array responses and the correlation between the effective communication and sensing channels. Favorable channel alignment allows a common reflection configuration to support both functions, consistent with the subspace rotation principle in \cite{S&C}. With multiple users, rotation additionally changes the users’ directions relative to the surface and their effective channel correlations. An orientation that improves the alignment between one user and the target may also increase inter-user channel correlation. To this end, it is challenging to jointly design the BS beamforming, SIRS orientation, and passive beamforming to minimize the CRB while satisfying the per-user communication constraints.

Motivated by the above considerations, we investigate a rotatable SIRS-assisted ISAC system, where a multi-antenna BS serves multiple single-antenna users and a point target in the non-line-of-sight (NLoS) region of the BS. In the considered system, the reflecting elements establish reflected links from the BS to the users and the target, while the sensors at the SIRS directly receive and process target echoes. Specifically, we consider a target-tracking scenario with prior target information, where the angle of the point target w.r.t. the SIRS is refined. By jointly adjusting the orientation of the reflecting panel and that of the sensors, SIRS rotation changes both the effective aperture gains and the array responses, thereby providing additional spatial DoFs for communication and sensing. To exploit these benefits, we aim to minimize the CRB for the target's spatial-frequency estimation by jointly optimizing the BS beamforming, SIRS beamforming, and rotation angles, subject to user SINR, transmit power, and reflection constraints. The rotation angles are nonlinearly coupled with the BS and SIRS beamformers, and the feasible orientations are restricted by the half-space reflection requirement, which requires joint beamforming and rotation designs. The main contributions are summarized as follows:
\begin{itemize}
    \item First, we propose angle-dependent communication and sensing models and derive the sensing CRB. For the single-path sensing-only and communication-only cases, we derive closed-form expressions for the beamforming and rotation designs, and show that two rotation angles are sufficient to achieve the performance of three-dimensional orientation under the considered setup. For a co-located user and target, we characterize the feasibility condition under which the sensing-optimal orientation also satisfies the user requirement.
    \item Second, we analyze the single-user case with unit effective aperture gains and characterize the impact of channel alignment on the performance of the ISAC system. We provide an SIRS orientation and beamforming design scheme to support communication and sensing, thereby distinguishing the spatial correlation benefit from the effective aperture gain. This analysis explains how rotation can balance the trade-off between the two functions through a proper reflection geometry. Furthermore, we develop an effective optimization framework for the multi-user case. A genetic algorithm is proposed for rotation optimization, while the BS and SIRS beamforming are optimized through semidefinite relaxation and Schur’s complement. 
    \item Finally, comprehensive numerical results are conducted to verify our analytical findings and demonstrate the effectiveness of the rotatable SIRS over the fixed-orientation SIRS. Moreover, the proposed rotation designs can achieve CRB performance close to that of exhaustive search and outperforms the communication-orientation scheme. The results further characterize the impact of transmit power, communication threshold, as well as the number of BS antennas, reflecting elements, and sensors on the ISAC performance.
\end{itemize}

The remainder of this paper is organized as follows. Section II presents the system model and the sensing CRB. Section III formulates the joint optimization problem, presents analytical designs for special cases, and develops effective methods for the single-user and multi-user scenarios. Section IV provides numerical results to evaluate the proposed designs and illustrate the trade-off between communication and sensing. Section V concludes the paper.

Notations: Scalars, vectors, and matrices are denoted by italic letters, bold lowercase letters, and bold uppercase letters, respectively. The transpose, complex conjugate, and conjugate transpose are represented by $(\cdot)^T$, $(\cdot)^*$, and $(\cdot)^H$, respectively. For a complex scalar $a$, $|a|$, $\arg(a)$, and $\operatorname{Re}\{a\}$ denote its magnitude, phase, and real part, respectively. For a vector $\mathbf{a}$, $\|\mathbf{a}\|$ and $[\mathbf{a}]_n$ denote its Euclidean norm and $n$-th entry, respectively. For a matrix $\mathbf{A}$, $[\mathbf{A}]_{m,n}$, $\operatorname{tr}(\mathbf{A})$, and $\operatorname{rank}(\mathbf{A})$ denote its $(m,n)$-th entry, trace, and rank, respectively. $\operatorname{diag}(\mathbf{a})$ represents a diagonal matrix whose diagonal entries are given by $\mathbf{a}$. $\mathbf{I}_N$ denotes the $N\times N$ identity matrix, and $\mathbf{0}$ denotes a zero vector or matrix of appropriate dimensions. $\mathcal{CN}(\bm{\mu},\bm{\Sigma})$ denotes the circularly symmetric complex Gaussian distribution with mean vector $\bm{\mu}$ and covariance matrix $\bm{\Sigma}$. The imaginary unit is denoted by $\jmath=\sqrt{-1}$. $\mathcal{O}(\cdot)$ is the big-O computational complexity notation.

\section{System Model}
\label{System Model}
\begin{figure}[t]
		\centering
		\includegraphics[width=0.35\textwidth]{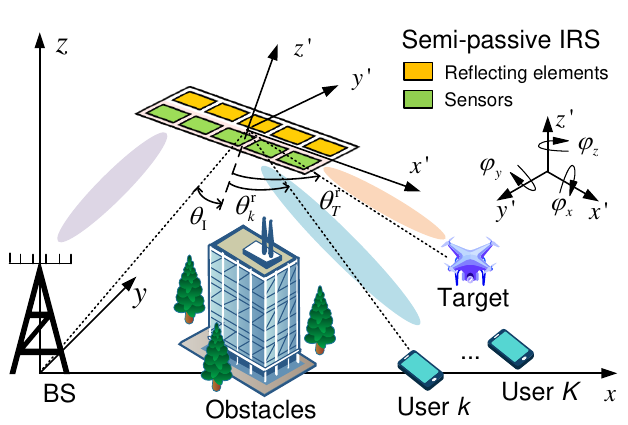}
		\vspace{-5pt}
		\caption{Illustration of a rotatable SIRS-aided ISAC system.}
		\label{fig:systemmodel}
        \vspace{-5pt}
\end{figure}

As shown in Fig. \ref{fig:systemmodel}, we consider a rotatable SIRS-aided ISAC system, where the SIRS is deployed to assist a multi-antenna BS to sense a point target and serve multiple single-antenna users. We focus on a challenging scenario where the direct BS-target and BS-user links are blocked by obstacles, and thus communication and target illumination rely on SIRS reflection. The SIRS and the BS are equipped with $N$ passive reflecting elements and $L$ active sensors, as well as $M$ transmit antennas, respectively. Let the set of the reflecting elements and users are denoted by $\mathcal{N} = \{1,\cdots, N\}$ and $\mathcal{K} = \{1,\cdots, K\}$, respectively, where the number of users is represented as $K$. In a global Cartesian coordinate system (CCS), the coordinates of the BS, SIRS, target, and user $k \in \mathcal{K}$ are given by ${{\mathbf{p}}_{\mathrm{B}}} = {\left[ {0,0,0} \right]^T}$, ${{\mathbf{p}}_{\mathrm{I}}} = {\left[ {{x_{\mathrm{I}}},{y_{\mathrm{I}}},{z_{\mathrm{I}}}} \right]^T}$, ${{\mathbf{p}}_{\mathrm{T}}} = {\left[ {{x_{\mathrm{T}}},{y_{\mathrm{T}}},{z_{\mathrm{T}}}} \right]^T}$, and ${{\mathbf{p}}_k} = {\left[ {{x_k},{y_k},{z_k}} \right]^T}$, respectively.

The 3D orientation of the SIRS can be represented by $\bm{\varphi} = \left[\varphi_x, \varphi_y, \varphi_z\right]^T$, where $\varphi_x \in \left[-\pi/2,\pi/2\right]$, $\varphi_y \in \left[-\pi/2,\pi/2\right]$, $\varphi_z \in \left[-\pi/2,\pi/2\right]$ are Euler angles denoting the SIRS’s degree of orientation around $x'$-, $y'$-, and $z'$-axis, respectively. The relationship between the global CCS and the local CCS is characterized by the orientation matrix expressed in \eqref{gCCS}.
\begin{figure*}[t]
	\begin{align}
		\label{gCCS}
		{\mathbf{Q}} ( \bm{\varphi} ) \!=\! \left[\!\!\! {\begin{array}{*{20}{c}}
				{{q_1}}&{{q_2}}&{{q_3}}\\
				{{q_4}}&{{q_5}}&{{q_6}}\\
				{{q_7}}&{{q_8}}&{{q_9}}
		\end{array}} \!\!\! \right] 
		\!=\! \left[\!\!\! {\begin{array}{*{20}{c}}
				{\cos {\varphi _y}\cos {\varphi _z}}&{\sin {\varphi _x}\sin {\varphi _y}\cos {\varphi _z} - \cos {\varphi _x}\sin {\varphi _z}}&{\cos {\varphi _x}\sin {\varphi _y}\cos {\varphi _z} + \sin {\varphi _x}\sin {\varphi _z}}\\
				{\cos {\varphi _y}\sin {\varphi _z}}&{\sin {\varphi _x}\sin {\varphi _y}\sin {\varphi _z} + \cos {\varphi _x}\cos {\varphi _z}}&{\cos {\varphi _x}\sin {\varphi _y}\sin {\varphi _z} - \sin {\varphi _x}\cos {\varphi _z}}\\
				{ - \sin {\varphi _y}}&{\sin {\varphi _x}\cos {\varphi _y}}&{\cos {\varphi _x}\cos {\varphi _y}}
		\end{array}} \!\!\! \right].
	\end{align}
	{\noindent} \rule[-0pt]{18.3cm}{0.05em}
\end{figure*}

Based on \eqref{gCCS}, we assume that the center of the SIRS is at the origin of the local CCS, i.e., ${\mathbf{p}}_{\mathrm{I}}^{\mathrm{L}} = {{\mathbf{Q}}^T}\left( {\bm{\varphi}} \right)\left( {{{\mathbf{p}}_{\mathrm{I}}} - {{\mathbf{p}}_{\mathrm{I}}}} \right) = {\left[ {0,0,0} \right]^T}$, and then the coordinates of the BS, SIRS, target, and user $k$ in the local CCS are given by	
\begin{align}
	\label{lCCS}
	{\mathbf{p}}_{\mathrm{B}}^{\mathrm{L}} = {{\mathbf{Q}}^T}\left( {\bm{\varphi }} \right)\left( {{{\mathbf{p}}_{\mathrm{B}}} - {{\mathbf{p}}_{\mathrm{I}}}} \right) = {\left[ {x_{\mathrm{B}}^{\mathrm{L}},y_{\mathrm{B}}^{\mathrm{L}},z_{\mathrm{B}}^{\mathrm{L}}} \right]^T}, \\
	{\mathbf{p}}_{\mathrm{T}}^{\mathrm{L}} = {{\mathbf{Q}}^T}\left( {\bm{\varphi }} \right)\left( {{{\mathbf{p}}_{\mathrm{T}}} - {{\mathbf{p}}_{\mathrm{I}}}} \right) = {\left[ {x_{\mathrm{T}}^{\mathrm{L}},y_{\mathrm{T}}^{\mathrm{L}},z_{\mathrm{T}}^{\mathrm{L}}} \right]^T}, \\
	{\mathbf{p}}_k^{\mathrm{L}} = {{\mathbf{Q}}^T}\left( {\bm{\varphi }} \right)\left( {{{\mathbf{p}}_k} - {{\mathbf{p}}_{\mathrm{I}}}} \right) = {\left[ {x_k^{\mathrm{L}},y_k^{\mathrm{L}},z_k^{\mathrm{L}}} \right]^T}.
\end{align}
Based on \eqref{lCCS}, the angle of departure (AoD) from the BS to the SIRS, the angle of arrival (AoA) at the SIRS from the BS, the AoD from the SIRS to user $k$, and the AoD from the SIRS to target, the AOA from target to the SIRS sensors are given by
\begin{align}
	& {\theta _{{\mathrm{BI}}}} = \arccos \frac{{{z_{\mathrm{I}}}}}{{\left\| {{{\mathbf{p}}_{\mathrm{I}}}} \right\|}}, \vartheta _{{\mathrm{IB}}} = \arccos \frac{{ - z_{\mathrm{B}}^{\mathrm{L}}}}{{\left\| {{{\mathbf{p}}_{\mathrm{I}}}} \right\|}}, \\
	& \theta _{\mathrm{I},k} \!=\! \arccos \frac{{z_k^{\mathrm{L}}}}{{\left\| {{{\mathbf{p}}_k} \!-\! {{\mathbf{p}}_{\mathrm{I}}}} \right\|}}, \theta _{{\mathrm{IT}}} = \vartheta_\mathrm{TI} \!=\! \arccos \frac{{z_{\mathrm{T}}^{\mathrm{L}}}}{{\left\| {{{\mathbf{p}}_{\mathrm{T}}} \!-\! {{\mathbf{p}}_{\mathrm{I}}}} \right\|}}.
\end{align}
The incident angle of the signal transmitted by the BS at the SIRS, the reflection angle w.r.t. user $k \in \mathcal{K}$, and the reflection angle w.r.t. the target are given by
\begin{align}
	{\theta _{\mathrm{I}}} =& \arccos \frac{{ - z_{\mathrm{B}}^{\mathrm{L}}}}{{\left\| {{{\mathbf{p}}_{\mathrm{I}}}} \right\|}},\\
	{\theta _k^\mathrm{r}} =& \arccos \frac{{ - z_k^{\mathrm{L}}}}{{\left\| {{{\mathbf{p}}_k} - {{\mathbf{p}}_{\mathrm{I}}}} \right\|}},\\
	{\theta _{\mathrm{T}}^\mathrm{r}} =& \arccos \frac{{ - z_{\mathrm{T}}^{\mathrm{L}}}}{{\left\| {{{\mathbf{p}}_{\mathrm{T}}} - {{\mathbf{p}}_{\mathrm{I}}}} \right\|}}.
\end{align}
To ensure that all users, the target, and the BS are located in the reflection space of the SIRS, we assume that ${\theta _{\mathrm{I}}} \in \left[0,\pi/2\right]$, ${\theta _k} \in \left[0,\pi/2\right]$, $k \in \mathcal{K}$, and ${\theta _{\mathrm{T}}} \in \left[0,\pi/2\right]$. Then, it follows that
\begin{align}
	\cos {\theta _{\mathrm{I}}} =& \frac{{ x_\mathrm{I} q_3  + y_\mathrm{I} q_6 + z_\mathrm{I} q_9 }}{{\sqrt {x_{\mathrm{I}}^2 + y_{\mathrm{I}}^2 + z_{\mathrm{I}}^2} }}, \label{theta_i}\\
	\cos {\theta _k^\mathrm{r}} =& \frac{{ ( {{x_{\mathrm{I}}} - {x_k}} )q_3  +  ( {{y_{\mathrm{I}}} - {y_k}} )q_6 +  ( {{z_{\mathrm{I}}} - {z_k}} ) q_9 }}{{\sqrt {{{\left( {{x_{\mathrm{I}}} - {x_k}} \right)}^2} + {{\left( {{y_{\mathrm{I}}} - {y_k}} \right)}^2} + {{\left( {{z_{\mathrm{I}}} - {z_k}} \right)}^2} } }},\\
	\cos {\theta _{\mathrm{T}}^\mathrm{r}} =& \frac{{\left( {{x_{\mathrm{I}}} - {x_{\mathrm{T}}}} \right)q_3 + \left( {{y_{\mathrm{I}}} - {y_{\mathrm{T}}}} \right)q_6 + \left( {{z_{\mathrm{I}}} - {z_{\mathrm{T}}}} \right)q_9 }}{{\sqrt {{{\left( {{x_{\mathrm{I}}} - {x_{\mathrm{T}}}} \right)}^2} + {{\left( {{y_{\mathrm{I}}} - {y_{\mathrm{T}}}} \right)}^2} + {{\left( {{z_{\mathrm{I}}} - {z_{\mathrm{T}}}} \right)}^2}} }}. \label{vartheta_TI}
\end{align}
As such, the effective aperture gain due to the SIRS’s orientation w.r.t. user $k$ and target can be expressed as
\begin{align}
	{F_k}\left( \bm{\varphi}  \right) = \cos {\theta _{\mathrm{I}}}\cos {\theta _k^{\mathrm{r}}},\\
		{F_{\mathrm{T}}}\left( \bm{\varphi}  \right) = \cos {\theta _{\mathrm{I}}}\cos {\theta _{\mathrm{T}}^{\mathrm{r}}}.
\end{align}

We consider an ISAC transmission block consisting of $T$ symbols, and the set of symbols is represented as $\mathcal{T} = \{1, \cdots , T\}$. The transmit signal at the BS is given by
\begin{align}
	{\mathbf{x}}\left( t \right) = \sum _{k \in \mathcal{K}}\mathbf{w}_{k}\mathbf{c}_{k}\left( t \right) + {\mathbf{s}}\left( t \right), t \in \mathcal{T},
\end{align}
where ${\mathbf{c}}(t) = \left[\mathbf{c}_{1}\left( t \right),\cdots,\mathbf{c}_{K}\left( t \right)\right]^T$ denotes the communication signal for users satisfying ${\mathbf{c}}\left( t \right) \sim \mathcal{CN}\left( {{\mathbf{0}},{{\mathbf{I}}_K}} \right)$, and ${\mathbf{s}}(t) \in \mathbb{C}^{M \times 1}$ represents the sensing signal with zero mean and covariance ${{\mathbf{R}}_{\mathrm{s}}} = \frac{1}{T}\sum\limits_{t = 1}^T {{\mathbf{s}}\left( t \right){{\mathbf{s}}^H}\left( t \right)} \succeq 0$. Assume that communication signals and sensing signals are statistically independent, the covariance matrix of the transmit signal is given by
\begin{align}
	{{\mathbf{R}}_{\mathrm{x}}} \!=\! \frac{1}{T}\sum\limits_{t = 1}^T {{\mathbf{x}}( t ){{\mathbf{x}}^H}( t )}  \approx \mathbb{E}\{ {{\mathbf{x}}\left( t \right){{\mathbf{x}}^H}( t )} \} \!=\! \sum _{k \in \mathcal{K}}\mathbf{W}_{k} \!+\! {{\mathbf{R}}_{\mathrm{s}}},
\end{align}
where $\mathbf{W}_{k} = \mathbf{w}_{k}\mathbf{w}_{k}^H$ with $\mathbf{W}_{k} \succeq 0$ and $\operatorname{rank} \left(\mathbf{W}_{k}\right) = 1$.
	
To facilitate the channel description, the steering vector function is defined as
\begin{align}
	{\mathbf{u}}\left( {\bar \theta ,\bar N} \right) = \left[ {{e^{ - \frac{{\jmath \pi \left( {\bar N - 1} \right)\bar \theta }}{2}}},{e^{ - \frac{{\jmath \pi \left( {\bar N - 3} \right)\bar \theta }}{2}}}, \cdots ,{e^{\frac{{\jmath \pi \left( {\bar N - 1} \right)\bar \theta }}{2}}}} \right],
\end{align}
where $\bar \theta$ denotes the phase difference between the observations at two adjacent elements and $\bar N$ is assumed as an even number denoting the size of an uniform linear array (ULA).
	
We assume far-field and quasi-static scenarios. The locations of the scatterers are assumed to be known a priori and remain fixed within each coherence block, which can be obtained from geometric priors or offline estimation. Let $C$ denote the number of scatterers. The coordinates of the $c$-th scatterer in the global CCS and local CCS are given by $\mathbf{p}_c = [x_c,y_c,z_c]^T$ and $\mathbf{p}_c^\mathrm{L} = [x_c^\mathrm{L},y_c^\mathrm{L},z_c^\mathrm{L}]^T$, $c \in \mathcal{C} = \{1, \cdots, C\}$, respectively. The BS-SIRS channel is modeled as
\begin{align}
    \label{path}
	{{\mathbf{H}}_{{\mathrm{BI}}}} = \sqrt{\frac{1}{C+1}} ({\bar {\mathbf{H}}_{{\mathrm{BI}}}} +  \sum\limits_{c = 1}^C {{{\tilde \alpha }_c}{{\bf{a}}_{{\rm{IB}}}}\left( {{\vartheta _c}} \right){\bf{a}}_{{\rm{BI}}}^T\left( {{\theta _c}} \right)} ),
\end{align}
where ${{\mathbf{a}}_{{\mathrm{IB}}}}\left( {\vartheta _{c}} \right) = {\mathbf{u}}\left( {\frac{{2{d_{\mathrm{I}}}}}{\lambda }\sin \vartheta _{c},N} \right)$, ${{\mathbf{a}}_{{\mathrm{BI}}}}\left( {{\theta _{c}}} \right) = {\mathbf{u}}\left( {\frac{{2{d_{\mathrm{B}}}}}{\lambda }\sin {\theta _{c}},M} \right)$ with ${\theta _c} = \arccos \frac{{{z_c}}}{{\left\| {{{\mathbf{p}}_c}} \right\|}}$ and $\vartheta _c = \arccos \frac{{ z_c^{\mathrm{L}}}}{{\left\| {{{\mathbf{p}}_c}} - {{{\mathbf{p}}_{\mathrm{I}}}} \right\|}}$, as well as ${{\tilde \alpha }_c} = \frac{\lambda^2}{(4\pi)^2 d_{\mathrm{B},c} d_{\mathrm{I},c}}
e^{-j\frac{2\pi}{\lambda}(d_{\mathrm{B},c} + d_{\mathrm{I},c})}$ denotes the path loss of the $c$-th path with $d_{\mathrm{B},c} = \|\mathbf{p}_c - \mathbf{p}_\mathrm{B}\|$ and $d_{\mathrm{I},c} = \|\mathbf{p}_\mathrm{I} - \mathbf{p}_c\|$. The SIRS elements spacing and the BS antenna spacing are denoted by $d_{\mathrm{I}}$ and $d_{\mathrm{B}}$, respectively, and $\lambda$ is the wavelength. The LoS-dominant path is given by
\begin{align}
	{\bar {\mathbf{H}}_{{\mathrm{BI}}}} = {\tilde {\alpha} _{{0}}}{{\mathbf{a}}_{{\mathrm{IB}}}}\left( \vartheta _{{\mathrm{IB}}} \right){\mathbf{a}}_{{\mathrm{BI}}}^T\left( {{\theta _{{\mathrm{BI}}}}} \right),
\end{align}
where ${\tilde {\alpha} _{{0}}} = \sqrt {\frac{{{\lambda ^2}}}{{16{\pi ^2}d_{{\mathrm{BI}}}^2}}} {e^{ - \jmath \frac{{2\pi {d_{{\mathrm{BI}}}}}}{\lambda }}}$ with $d_\mathrm{BI} = \|\mathbf{p}_\mathrm{B} - \mathbf{p}_\mathrm{I}\|$, ${{\mathbf{a}}_{{\mathrm{IB}}}}\left( {\vartheta _{{\mathrm{IB}}}} \right) = {\mathbf{u}}\left( {\frac{{2{d_{\mathrm{I}}}}}{\lambda }\sin \vartheta _{{\mathrm{IB}}},N} \right)$, and ${{\mathbf{a}}_{{\mathrm{BI}}}}\left( {{\theta _{{\mathrm{BI}}}}} \right) = {\mathbf{u}}\left( {\frac{{2{d_{\mathrm{B}}}}}{\lambda }\sin {\theta _{{\mathrm{BI}}}},M} \right)$.

\subsection{Communication Model}
The channel between SIRS and the user $k$ is modeled as
\begin{align}
	{\mathbf{h}}_{k}^H = {\alpha _k}{\mathbf{a}}_{k}^T\left( \theta _{\mathrm{I},k} \right),
\end{align}	
where ${\alpha _k} = \sqrt {\frac{{{\lambda ^2}}}{{16{\pi ^2}d_k^2}}} {e^{ - \jmath \frac{{2\pi {d_k}}}{\lambda }}}$ with $d_k = \|\mathbf{p}_k - \mathbf{p}_\mathrm{I}\|$, and ${{\mathbf{a}}_k}\left( \theta _{\mathrm{I},k} \right) = {\mathbf{u}}\left( {\frac{{2{d_{\mathrm{I}}}}}{\lambda }\sin \theta _{\mathrm{I},k},{N}} \right) $.

The signal received at the user $k$ is given by
\begin{align}
	{y_k}\left(t\right) = \sqrt{{F_k}\left( {\bm{\varphi }} \right)} {\mathbf{h}}_k^H{\mathbf{\Phi }}{{\mathbf{H}}_{{\mathrm{BI}}}}{\mathbf{x}}\left( t \right) + {n_k}\left(t\right),
\end{align}
where ${\mathbf{\Phi}} = \operatorname{diag}\left(e^{j\phi_1},\cdots,e^{j\phi_N} \right)$ denotes the reflection matrix of the SIRS and ${n_k}\left(t\right) \sim \mathcal{CN} \left(0, \sigma_k^2\right)$ denotes the additive white Gaussian noise (AWGN) at $k$-th user. We consider the user receivers are dedicated designed for ISAC, which can cancel the interference from the sensing signal. As such, the received SINR of user $k$ is given by
\begin{align}
	{\gamma _k} (\bm{\varphi}, \mathbf{W}_{k},\mathbf{\Phi} ) \!=\! \frac{{{F_k}\left( {\bm{\varphi }} \right)}{{{ {{\mathbf{h}}_{k}^H{\mathbf{\Phi }}{{\mathbf{H}}_{{\mathrm{BI}}}}{{\mathbf{W}}_k}}{\mathbf{H}}_{{\mathrm{BI}}}^H \mathbf{\Phi }^H {\mathbf{h}}_{k}}}}}{{\sum \limits_{i \in \mathcal{K}, i \ne k} \!\!\!\!\! {{F_k}\left( {\bm{\varphi }} \right)}{ {{{ {{\mathbf{h}}_{k}^H{\mathbf{\Phi }}{{\mathbf{H}}_{{\mathrm{BI}}}}{{\mathbf{W}}_i}}{\mathbf{H}}_{{\mathrm{BI}}}^H \mathbf{\Phi }^H {\mathbf{h}}_{k}}}} \!+\! \sigma _k^2} }}.
\end{align}

\subsection{Sensing Model}
The SIRS element-target-SIRS sensor channel is given by
\begin{align}
	{{\mathbf{H}}_{{\mathrm{ITI}}}} = {\alpha _{{\mathrm{ITI}}}}{{\mathbf{a}}_{{\mathrm{TI}}}}\left( {{\vartheta _{{\mathrm{TI}}}}} \right){\mathbf{a}}_{{\mathrm{IT}}}^T\left( \theta _{{\mathrm{IT}}} \right),
\end{align}
where ${\alpha _{{\mathrm{ITI}}}} = {\beta _0}{\beta _{{\mathrm{ITI}}}}$ with ${\beta _0} \sim \mathcal{CN}\left( {0,1} \right)$ and $\beta _{{\mathrm{ITI}}} = \sqrt{\frac{\lambda^2 \kappa}{64 \pi^3 d^4_\mathrm{IT}}}$, ${{\mathbf{a}}_{{\mathrm{TI}}}}\left( {{\vartheta _{{\mathrm{TI}}}}} \right) = {\mathbf{u}}\left( {\frac{{2{d_{\mathrm{S}}}}}{\lambda }\sin {\vartheta _{{\mathrm{TI}}}},L} \right)$ with the SIRS sensor spacing $d_{\mathrm{S}}$, and ${{\mathbf{a}}_{{\mathrm{IT}}}}\left( \theta _{{\mathrm{IT}}} \right) = {\mathbf{u}}\left( {\frac{{2{d_{\mathrm{I}}}}}{\lambda }\sin \theta _{{\mathrm{IT}}},{N}} \right)$. The distance between the SIRS and the target is denoted by $d_\mathrm{IT} = \|\mathbf{p}_\mathrm{T} - \mathbf{p}_\mathrm{I}\|$ and $\kappa$ represents the radar cross-section. 
	
It is assumed that the direct-path component from the BS to the sensor can be acquired a priori via offline estimation and perfectly canceled at the sensor before target's spatial-frequency estimation. Thus, the received echo signal at time symbol $t$ is given by 
\begin{align}
	\mathbf{y}_{\mathrm{s}}(t) = \sqrt{F_\mathrm{T}(\bm{\varphi})} {{\mathbf{H}}_{{\mathrm{ITI}}}} \mathbf{\Phi} {{\mathbf{H}}_{{\mathrm{BI}}}} \mathbf{x} (t) +  \mathbf{n}_{\mathrm{T}} (t),
\end{align}
where $\mathbf{n}_{\mathrm{T}} (t) \sim \mathcal{CN} \left(0, \sigma _\mathrm{T}^2 \mathbf{I}_{L} \right)$ denotes the received noise at the SIRS sensors.
	
Let $\bar \vartheta _{{\mathrm{TI}}} = \sin \vartheta _{{\mathrm{TI}}} = \sin \theta _ {\mathrm{IT}}$ and ${\mathbf{A}} = {{\mathbf{a}}_{{\mathrm{TI}}}}( \bar \vartheta _{{\mathrm{TI}}} ){\mathbf{a}}_{{\mathrm{IT}}}^T( \bar \vartheta _{{\mathrm{TI}}} ) \mathbf{\Phi} {{\mathbf{H}}_{{\mathrm{BI}}}} $. The CRB of the estimated $\bar \vartheta _{{\mathrm{TI}}}$ is given by
\begin{align}
	\operatorname{CRB}{\!( \bar \vartheta _{{\mathrm{TI}}} \!)} \!=\! \frac{{\sigma _{\mathrm{T}}^2}}{{2T{{| {{\alpha _{{\mathrm{ITI}}}}} |}^2} {F_\mathrm{T}(\bm{\varphi})} ( {\operatorname{tr} ( {{\mathbf{\dot{A}}}{{\mathbf{R}}_{\mathrm{x}}}{{{\mathbf{\dot{A}}}}^H}} ) \!-\! \frac{{{{| {\operatorname{tr} ( {{\mathbf{A}}{{\mathbf{R}}_{\mathrm{x}}}{{{\mathbf{\dot{A}}}}^H}} )} |}^2}}}{{\operatorname{tr} ( {{\mathbf{A}}{{\mathbf{R}}_{\mathrm{x}}}{{\mathbf{A}}^H}} )}}} )}}.
\end{align}

Assuming that the inter-element distance and the inter-sensor distance are the same, i.e., $d_\mathrm{s} = d_\mathrm{I} = \bar d$, the derivative of ${{\mathbf{a}}_{{\mathrm{TI}}}}\left( \bar \vartheta _{{\mathrm{TI}}} \right)$ and ${{\mathbf{a}}_{{\mathrm{IT}}}}\left( \bar \vartheta _{{\mathrm{TI}}} \right)$ are given by 
\begin{align}
	{{\mathbf{\dot a}}_{{\mathrm{TI}}}}\left( \bar \vartheta _{{\mathrm{TI}}} \right) &= \frac{{\partial {{\mathbf{ a}}_{{\mathrm{TI}}}}\left( \bar \vartheta _{{\mathrm{TI}}} \right)}}{{\partial \bar \vartheta _{{\mathrm{TI}}}}}= j \pi \frac{\bar d}{\lambda}{{\mathbf{D}}_{\mathrm{TI}}} {{\mathbf{ a}}_{{\mathrm{TI}}}}\left( \bar \vartheta _{{\mathrm{TI}}} \right), \label{adot}\\
	{{\mathbf{\dot a}}_{{\mathrm{IT}}}}\left( \bar \vartheta _{{\mathrm{TI}}} \right) &= \frac{{\partial {{\mathbf{ a}}_{{\mathrm{IT}}}}\left( \bar \vartheta _{{\mathrm{TI}}} \right)}}{{\partial \bar \vartheta _{{\mathrm{TI}}} }} = j \pi \frac{\bar d}{\lambda}{{\mathbf{D}}_{\mathrm{IT}}}{{\mathbf{ a}}_{{\mathrm{IT}}}}\left( \bar \vartheta _{{\mathrm{TI}}} \right), \label{bdot}
\end{align}
with ${{\mathbf{D}}_{\mathrm{TI}}} = {\operatorname{diag}} ( { - ( {L - 1} ), - ( {L - 3} ), \ldots , ( {L - 1} )} )$ and ${{\mathbf{D}}_{\mathrm{IT}}} = {\operatorname{diag}} ( { - ( {N - 1} ), - ( {N - 3} ), \ldots , ( {N - 1} )} )$. Let $\mathbf v=[e^{j\phi_1},\ldots,e^{j\phi_N}]^T$ denote the IRS reflection coefficient vector, such that $\bm\Phi=\operatorname{diag}(\mathbf v)$ and $|v_n|=1$, $\forall n\in\mathcal N$. By introducing ${\mathbf{B}} = \operatorname{diag} \left({\mathbf{a}}_{{\mathrm{IT}}}\left( \bar \vartheta _{{\mathrm{TI}}} \right)\right)$, we have 
\begin{align}
	&{\mathbf{A}} = {{\mathbf{a}}_{{\mathrm{TI}}}}( \bar \vartheta _{{\mathrm{TI}}} ){\mathbf{a}}_{{\mathrm{IT}}}^T( \bar \vartheta _{{\mathrm{TI}}} ) \mathbf{\Phi} {{\mathbf{H}}_{{\mathrm{BI}}}}  = {{\mathbf{a}}_{{\mathrm{TI}}}}( \bar \vartheta _{{\mathrm{TI}}} ) \mathbf{v}^T \mathbf{B} {{\mathbf{H}}_{{\mathrm{BI}}}}, \label{A}\\
	&\mathbf{\dot A} = {{\mathbf{\dot a}}_{{\mathrm{TI}}}}( \bar \vartheta _{{\mathrm{TI}}} ){\mathbf{a}}_{{\mathrm{IT}}}^T( \bar \vartheta _{{\mathrm{TI}}} ) \mathbf{\Phi} {{\mathbf{H}}_{{\mathrm{BI}}}} + {{\mathbf{a}}_{{\mathrm{TI}}}}( \bar \vartheta _{{\mathrm{TI}}} ){\mathbf{\dot a}}_{{\mathrm{IT}}}^T( \bar \vartheta _{{\mathrm{TI}}} ) \mathbf{\Phi} {{\mathbf{H}}_{{\mathrm{BI}}}} \nonumber\\
	&= j \pi \frac{{{\bar d}}}{{{\lambda}}} ({{\mathbf{D}}_{\mathrm{TI}}} {{\mathbf{a}}_{{\mathrm{TI}}}}( \bar \vartheta _{{\mathrm{TI}}} ) \mathbf{v}^T \mathbf{B} {{\mathbf{H}}_{{\mathrm{BI}}}} + {{\mathbf{a}}_{{\mathrm{TI}}}}{{\mathbf{v}}^T}{\mathbf{D}_\mathrm{IT}}{\mathbf{B {{\mathbf{H}}_{{\mathrm{BI}}}}}}). \label{Adot}
\end{align}
Based on \eqref{adot}, \eqref{bdot}, \eqref{A} and \eqref{Adot}, it yields
\begin{align}
	{\operatorname{tr}} ( {{\mathbf{A}}{{\mathbf{R}}_{\mathrm{x}}}{{\mathbf{A}}^H}} ) 
	= & L {{{\mathbf{v}}^H}{{\mathbf{R}}_1}{\mathbf{v}}}, \label{BB} \\
	{\operatorname{tr}} ( {{\mathbf{A}}{{\mathbf{R}}_{\mathrm{x}}}{{{\mathbf{\dot A}}}^H}} )
	= & -j \pi \frac{{{\bar d}}}{{{\lambda}}} L {{{\mathbf{v}}^H}{\mathbf{D}_\mathrm{IT}}{{\mathbf{R}}_1}{\mathbf{v}}}, \label{BBdot} \\
	{\operatorname{tr}} ( {{\mathbf{\dot A}}{{\mathbf{R}}_{\mathrm{x}}}{{{\mathbf{\dot A}}}^H}} ) 
	=&  {\pi ^2}\frac{{{\bar d}^2}}{{\lambda ^2}} \frac{L^3-L}{3} {{{\mathbf{v}}^H}{{\mathbf{R}}_1}{\mathbf{v}}} \nonumber\\
	&+ {\pi ^2}\frac{{{\bar d}^2}}{{\lambda ^2}} L {{{\mathbf{v}}^H}{\mathbf{D}_\mathrm{IT}}{{\mathbf{R}}_1}{\mathbf{D}_\mathrm{IT}}{\mathbf{v}}}, \label{BdotBdot}
\end{align}
with ${{\mathbf{R}}_1} = {{\mathbf{B}^*}}{{\mathbf{H}}_{{\mathrm{BI}}}^*}{{\mathbf{R}}^T_{\mathrm{x}}}{{\mathbf{H}}_{{\mathrm{BI}}}^T}{\mathbf{B}}$.
As such, the CRB can be rewritten as \eqref{CRB} on the top of next page.
\begin{figure*}[t]
	\begin{align}
		\label{CRB}
		{\operatorname{CRB}}( \bar \vartheta _{{\mathrm{TI}}} ) = \frac{{\sigma _{\mathrm{T}}^2\lambda ^2}}{{2T{{\left| {{\alpha _{{\mathrm{ITI}}}}} \right|}^2}{\pi ^2}{{\bar d}^2} F_\mathrm{T} \left(\bm{\varphi}\right)\left( {( {\frac{{{L^3} - L}}{3}} ) ( {{{\mathbf{v}}^H}{{\mathbf{R}}_1}{\mathbf{v}}} ) + L ( {{{\mathbf{v}}^H}{{\mathbf{D}}_{\mathrm{IT}}}{{\mathbf{R}}_1}{{\mathbf{D}}_{\mathrm{IT}}}{\mathbf{v}}} ) - L\frac{{{{\left| {{{\mathbf{v}}^H}{\mathbf{D}_\mathrm{IT}}{{\mathbf{R}}_1}{\mathbf{v}}} \right|}^2}}}{{( {{{\mathbf{v}}^H}{{\mathbf{R}}_1}{\mathbf{v}}} )}}} \right)}}.
	\end{align}
	{\noindent} \rule[-0pt]{18.3cm}{0.05em}
\end{figure*}

\section{Problem Formulation and Solutions}
In this section, we first formulate a joint optimization problem to minimize the CRB for target's spatial-frequency estimation while satisfying the SINR threshold of the communication users. Then, we analyze special cases to characterize the impact of SIRS rotation on the effective aperture gains and channel correlation, which provides useful insights and lays the foundation for the general multi-user case.

Based on the derived CRB and the angle-dependent model, we aim to minimize the CRB for target's spatial-frequency estimation by jointly optimizing the BS beamforming, the SIRS beamforming, and the rotation angles. Thus, the SINR-constrained CRB minimization problem is formulated as
\begin{subequations}
	\label{pro:original}
	\begin{align}
		\mathop {\min }\limits_{\substack{\mathbf{\Phi}, \{\mathbf{W}_{k}\}, \\ \mathbf{R}_\mathrm{s}, \bm{\varphi}}} &\operatorname{CRB}{\left( \bar \vartheta _{{\mathrm{TI}}} \right)} \\
		\mathrm{s.t.} \;\;\; & {\gamma _k} \left(\bm{\varphi}, \mathbf{W}_{k},\mathbf{\Phi} \right) \ge \Gamma_k, \forall k \in \mathcal{K}, \label{sinr}\\
		& \sum_{k \in \mathcal{K}}\operatorname{tr}\left(\mathbf{W}_{k}\right) + \operatorname{tr}\left(\mathbf{R}_\mathrm{s}\right) \le P_0, \label{txpower}\\
		& \mathbf{R}_\mathrm{s} \succeq 0, \label{Rs} \\
		&\mathbf{W}_{k} \succeq 0, \forall k \in \mathcal{K}, \label{wk}\\
		& \operatorname{rank} (\mathbf{W}_{k}) = 1, \forall k \in \mathcal{K}, \label{rankwk} \\
		& {x_{\mathrm{I}}}{q_3} \!+\! {y_{\mathrm{I}}}{q_6} \!+\! {z_{\mathrm{I}}}{q_9} \ge 0, \label{BSrotation}\\
		& ( {{x_{\mathrm{I}}} \!-\! {x_{\mathrm{T}}}} ){q_3} + ( {{y_{\mathrm{I}}} \!-\! {y_{\mathrm{T}}}} ){q_6} \!+\! ( {{z_{\mathrm{I}}} \!- \! {z_{\mathrm{T}}}} ){q_9} \ge 0, \label{targetrotation}\\
		& ( {{x_{\mathrm{I}}} \!-\! {x_k}} ){q_3} \!+\! ( {{y_{\mathrm{I}}} \!-\! {y_k}} ){q_6} \!+\! ( {{z_{\mathrm{I}}} \!-\! {z_k}} ){q_9} \ge 0, \forall k \in \mathcal{K}, \label{userrotation} \\
		&|{\mathbf{\Phi}}_{n,n}|=1, \forall n \in \mathcal{N}, \label{rank1}
	\end{align}
\end{subequations}
where $P_0$ denote the maximum tansmit power at the BS, $\Gamma_k, \forall k \in \mathcal{K}$ represents the pre-defined communication SINR threshold w.r.t. user $k$. For problem \eqref{pro:original}, constraints \eqref{sinr} and \eqref{txpower} are the per-user SINR requirement and the maximum power budget, respectively, while constraints \eqref{Rs} and \eqref{wk} are the positive semidefinite constraints on the optimization variables. Constraint \eqref{rankwk} imposes the rank-one condition. Constraints \eqref{BSrotation}–\eqref{userrotation} ensure that the BS, the target, and all users lie in the reflection half-space of the SIRS. Constraint \eqref{rank1} is the unit-modulus constraint on each IRS phase shift. Due to the coupling between the rotation-dependent aperture gains, the BS beamformer, and the IRS beamformer, problem \eqref{pro:original} is non-convex and thus challenging to solve.
    
In the following, we discuss special cases under the BS–SIRS channel model with only a LoS-dominant path. Specifically, for the sensing-only and communication-only cases, the CRB for estimating $\bar \vartheta_{\mathrm{TI}}$ and the communication SNR are minimized and maximized, respectively. We then consider a co-located user and target case to determine the feasibility condition imposed by the communication requirement. In the case with a separated user and target, we design the SIRS rotation to improve the correlation between communication and sensing. For clarity, we denote the user index as U. 
	
\subsection{Sensing-Only Case}
We first consider the special case of target sensing, and then we have ${\mathbf{A}} = {\tilde {\alpha} _{{0}}} {{\mathbf{a}}_{{\mathrm{TI}}}}{\left( {{\bar \vartheta _{{\mathrm{TI}}}}} \right)}{\mathbf{a}}_{{\mathrm{IT}}}^T{\left( {{\bar \vartheta _{{\mathrm{TI}}}}} \right)}{\mathbf{\Phi }}{{\mathbf{a}}_{{\mathrm{IB}}}}\left( {{\vartheta _{{\mathrm{IB}}}}} \right){\mathbf{a}}_{{\mathrm{BI}}}^T\left( {{\theta _{{\mathrm{BI}}}}} \right)$ and $\mathbf{\dot A} = {\tilde {\alpha} _{{0}}} {{\mathbf{\dot a}}_{{\mathrm{TI}}}}( \bar \vartheta _{{\mathrm{TI}}} ){\mathbf{a}}_{{\mathrm{IT}}}^T( \bar \vartheta _{{\mathrm{TI}}} ) \mathbf{\Phi} {{\mathbf{a}}_{{\mathrm{IB}}}}\left( {{\vartheta _{{\mathrm{IB}}}}} \right){\mathbf{a}}_{{\mathrm{BI}}}^T\left( {{\theta _{{\mathrm{BI}}}}} \right) +{\tilde {\alpha} _{{0}}} {{\mathbf{a}}_{{\mathrm{TI}}}}( \bar \vartheta _{{\mathrm{TI}}} ){\mathbf{\dot a}}_{{\mathrm{IT}}}^T( \bar \vartheta _{{\mathrm{TI}}} ) \mathbf{\Phi} {{\mathbf{a}}_{{\mathrm{IB}}}}\left( {{\vartheta _{{\mathrm{IB}}}}} \right){\mathbf{a}}_{{\mathrm{BI}}}^T\left( {{\theta _{{\mathrm{BI}}}}} \right)$, which yields
\begin{align}
	&\operatorname{tr} ( {{\mathbf{A}}{{\mathbf{R}}_{\mathrm{x}}}{{\mathbf{A}}^H}} ) = {|\alpha _{{\mathrm{BI}}}|^2}{\left\| {{{\mathbf{a}}_{{\mathrm{TI}}}}{\left( {{\bar \vartheta _{{\mathrm{TI}}}}} \right)}} \right\|^2}{\left| {{\mathbf{a}}_{{\mathrm{IT}}}^T{\left( {{\bar \vartheta _{{\mathrm{TI}}}}} \right)}{\mathbf{\Phi }}{{\mathbf{a}}_{{\mathrm{IB}}}}\left( {{\vartheta _{{\mathrm{IB}}}}} \right)} \right|^2}\nonumber\\
	&\times {\mathbf{a}}_{{\mathrm{BI}}}^T\left( {{\theta _{{\mathrm{BI}}}}} \right){{\mathbf{R}}_{\mathrm{x}}}{{\mathbf{a}}_{{\mathrm{BI}}}^*}{\left( {{\theta _{{\mathrm{BI}}}}} \right)}, \label{A1} \\
	&\operatorname{tr} ( {{\mathbf{A}}{{\mathbf{R}}_{\mathrm{x}}}{{{\mathbf{\dot{A}}}}^H}} ) = {|\alpha _{{\mathrm{BI}}}|^2} {\left\| {{{\mathbf{a}}_{{\mathrm{TI}}}}{\left( {{\bar \vartheta _{{\mathrm{TI}}}}} \right)}} \right\|^2}{\mathbf{a}}_{{\mathrm{IT}}}^T{\left( {{\bar \vartheta _{{\mathrm{TI}}}}} \right)}{\mathbf{\Phi }}{{\mathbf{a}}_{{\mathrm{IB}}}}\left( {{\vartheta _{{\mathrm{IB}}}}} \right)\nonumber\\
	&\times {\mathbf{a}}_{{\mathrm{BI}}}^T\left( {{\theta _{{\mathrm{BI}}}}} \right){{\mathbf{R}}_{\mathrm{x}}}{{\mathbf{a}}^*_{{\mathrm{BI}}}}\left( {{\theta _{{\mathrm{BI}}}}} \right) {\mathbf{a}}_{{\mathrm{IB}}}^H\left( {{\vartheta _{{\mathrm{IB}}}}} \right){{\mathbf{\Phi }}^H}{{{\mathbf{\dot a}}}^*_{{\mathrm{IT}}}}{\left( {{\bar \vartheta _{{\mathrm{TI}}}}} \right)},  \label{A2} \\
	&\operatorname{tr} ( {{\mathbf{\dot{A}}}{{\mathbf{R}}_{\mathrm{x}}}{{{\mathbf{\dot{A}}}}^H}} ) = {|\alpha _{{\mathrm{BI}}}|^2} {\left\| {{{{\mathbf{\dot a}}}_{{\mathrm{TI}}}}{\left( {{\bar \vartheta _{{\mathrm{TI}}}}} \right)}} \right\|^2}{\left| {{\mathbf{a}}_{{\mathrm{IT}}}^T{\left( {{\bar \vartheta _{{\mathrm{TI}}}}} \right)}{\mathbf{\Phi }}{{\mathbf{a}}_{{\mathrm{IB}}}}\left( {{\vartheta _{{\mathrm{IB}}}}} \right)} \right|^2}\nonumber\\
	&\times {\mathbf{a}}_{{\mathrm{BI}}}^T\left( {{\theta _{{\mathrm{BI}}}}} \right){{\mathbf{R}}_{\mathrm{x}}}{{\mathbf{a}}^*_{{\mathrm{BI}}}}\left( {{\theta _{{\mathrm{BI}}}}} \right) + {|\alpha _{{\mathrm{BI}}}|^2} {\left\| {{{\mathbf{a}}_{{\mathrm{TI}}}}{\left( {{\bar \vartheta _{{\mathrm{TI}}}}} \right)}} \right\|^2}\nonumber\\
	&\times {\left| {{\mathbf{\dot a}}_{{\mathrm{IT}}}^T{\left( {{\bar \vartheta _{{\mathrm{TI}}}}} \right)}{\mathbf{\Phi }}{{\mathbf{a}}_{{\mathrm{IB}}}}\left( {{\vartheta _{{\mathrm{IB}}}}} \right)} \right|^2}{\mathbf{a}}_{{\mathrm{BI}}}^T\left( {{\theta _{{\mathrm{BI}}}}} \right){{\mathbf{R}}_{\mathrm{x}}}{{\mathbf{a}}^*_{{\mathrm{BI}}}}\left( {{\theta _{{\mathrm{BI}}}}} \right). \label{A3}
\end{align}
Then, it follows that
\begin{align}
&{| {\operatorname{tr} ( {{\mathbf{A}}{{\mathbf{R}}_{\mathrm{x}}}{{{\mathbf{\dot A}}}^H}} )} |}^2 
= |{\tilde {\alpha} _{{0}}}|^4{\| {{{\mathbf{a}}_{{\mathrm{TI}}}}( {{\bar \vartheta _{{\mathrm{TI}}}}} )} \|^4} {| {{\mathbf{\dot a}}_{{\mathrm{IT}}}^T( {{\bar \vartheta _{{\mathrm{TI}}}}} ){\mathbf{\Phi }}{{\mathbf{a}}_{{\mathrm{IB}}}}( {{\vartheta _{{\mathrm{IB}}}}} )} |^2} \nonumber \\
&\times {| {{\mathbf{a}}_{{\mathrm{IT}}}^T( {{\bar \vartheta _{{\mathrm{TI}}}}} ){{\mathbf{\Phi }}}{{\mathbf{a}}_{{\mathrm{IB}}}}( {{\vartheta _{{\mathrm{IB}}}}} )} |^2} | {{\mathbf{a}}_{{\mathrm{BI}}}^T( {{\theta _{{\mathrm{BI}}}}} ){{\mathbf{R}}_{\mathrm{x}}}{{\mathbf{a}}^*_{{\mathrm{BI}}}}( {{\theta _{{\mathrm{BI}}}}} )} |^2. \label{A22}
\end{align}
Based on \eqref{A1} and \eqref{A22}, we have $\frac{{{{| {\operatorname{tr} ( {{\mathbf{A}}{{\mathbf{R}}_{\mathrm{x}}}{{{\mathbf{\dot{A}}}}^H}} )} |}^2}}}{{\operatorname{tr} ( {{\mathbf{A}}{{\mathbf{R}}_{\mathrm{x}}}{{\mathbf{A}}^H}} )}} =  {|\alpha _{{\mathrm{BI}}}|^2} {\| {{{\mathbf{a}}_{{\mathrm{TI}}}}{( {{\bar \vartheta _{{\mathrm{TI}}}}} )}} \|^2} {| {{\mathbf{\dot a}}_{{\mathrm{IT}}}^T{( {{\bar \vartheta _{{\mathrm{TI}}}}} )}{\mathbf{\Phi }}{{\mathbf{a}}_{{\mathrm{IB}}}}( {{\vartheta _{{\mathrm{IB}}}}} )} |^2}{\mathbf{a}}_{{\mathrm{BI}}}^T( {{\theta _{{\mathrm{BI}}}}} ){{\mathbf{R}}_{\mathrm{x}}}{{\mathbf{a}}^*_{{\mathrm{BI}}}}( {{\theta _{{\mathrm{BI}}}}} )$. As such, the CRB of the estimated ${{\bar \vartheta _{{\mathrm{TI}}}}}$ is expressed in \eqref{CRB_LoS}.
\begin{figure*}
	\begin{align}
		\label{CRB_LoS}
		\operatorname{CRB}{\left( {{\bar \vartheta _{{\mathrm{TI}}}}} \right)} = \frac{3 {\sigma _{\mathrm{T}}^2} \lambda ^2}{{2T{|\alpha _{{\mathrm{BI}}}|^2}{\left| {{\alpha _{{\mathrm{ITI}}}}} \right|^2} \pi^2 \bar{d}^2 (L^3 - L) {{F_{\mathrm{T}}}\left( \bm{\varphi}  \right) {\left| {{\mathbf{a}}_{{\mathrm{IT}}}^T{\left( \bar \vartheta _{{\mathrm{TI}}} \right)}{\mathbf{\Phi }}{{\mathbf{a}}_{{\mathrm{IB}}}}\left( {{\vartheta _{{\mathrm{IB}}}}} \right)} \right|^2}{\mathbf{a}}_{{\mathrm{BI}}}^T\left( {{\theta _{{\mathrm{BI}}}}} \right){{\mathbf{R}}_{\mathrm{x}}}{{\mathbf{a}}^*_{{\mathrm{BI}}}}\left( {{\theta _{{\mathrm{BI}}}}} \right)} }}.
	\end{align}
	{\noindent} \rule[-0pt]{18.3cm}{0.05em}
\end{figure*}

Based on \eqref{CRB_LoS}, problem \eqref{pro:original} is reduced to
\begin{subequations}
	\label{pro:CRB}
	\begin{align}
		\mathop {\min }\limits_{\mathbf{\Phi},\mathbf{R}_\mathrm{s},\bm{\varphi}} \; &\operatorname{CRB}{\left( \bar \vartheta _{{\mathrm{TI}}} \right)} \\
		\mathrm{s.t.} \;\;\;\; 
		& \operatorname{tr}\left(\mathbf{R}_\mathrm{s}\right) \le P_0,\\
		& \eqref{Rs}, \eqref{BSrotation}, \eqref{targetrotation}, \eqref{rank1}.
	\end{align}
\end{subequations}
	
Given $\mathbf{\Phi}$ and $\bm{\varphi}$, the optimal beamforming at the BS is given by
\begin{align}
	\label{BS_BF_S}
	{\mathbf{R}}_{\mathrm{s}}^{\star} = {P_0}\frac{{{{\mathbf{a}}^*_{{\mathrm{BI}}}}\left( {{\theta _{{\mathrm{BI}}}}} \right){\mathbf{a}}_{{\mathrm{BI}}}^T\left( {{\theta _{{\mathrm{BI}}}}} \right)}}{{\left\| {{{\mathbf{a}}_{{\mathrm{BI}}}}\left( {{\theta _{{\mathrm{BI}}}}} \right)} \right\|^2}}.
\end{align}
	
Given $\mathbf{R}_\mathrm{s}$ and $\bm{\varphi}$, the optimal SIRS phase shift is given by
\begin{align}
	\label{IRS_BF_S}
		\phi_n^{\star} = -\arg \left[{{{\mathbf{a}}_{{\mathrm{IT}}}}\left( {{\vartheta _{{\mathrm{TI}}}}} \right)} \right]_n  - \arg \left[ {{{\mathbf{a}}_{{\mathrm{IB}}}}\left( {{\vartheta _{{\mathrm{IB}}}}} \right)} \right]_n , \forall n \in \mathcal{N}.
\end{align}
	
Given $\mathbf{\Phi}^{\star}$ and $\mathbf{R}_\mathrm{s}^{\star}$, the CRB w.r.t. $\bm{\varphi}$ is given by
\begin{align}
	\operatorname{CRB} (\bm{\varphi}) \!=\!  \frac{{3\sigma _{\mathrm{T}}^2 {\lambda ^2} }}{{2T{|\alpha _{{\mathrm{BI}}}|^2}{\left| {{\alpha _{{\mathrm{ITI}}}}} \right|^2}{\pi ^2} {\bar d^2} {P_0}M{N^2}( {{L^3} - L} ){F_{\mathrm{T}}}( \bm{\varphi}  ) }}.
	\label{sensing_crb}
\end{align}

\begin{Proposition}
	\label{pro:2D}
	For any given SIRS's 3D orientation $\bm{\varphi}_\mathrm{3D} = \left[\varphi_x, \varphi_y, \varphi_z\right]^T$ and location, there always exists a 2D orientation solution $\bm{\varphi}_\mathrm{2D}^{\star} = \left[\varphi_x^{\star}, \varphi_y^{\star}, 0\right]^T$ satisfying $\operatorname{CRB} \left(\bm{\varphi}_\mathrm{2D}^{\star}\right) = \operatorname{CRB} \left(\bm{\varphi}_\mathrm{3D}\right)$.
\end{Proposition}
{\it{Proof:}} Please refer to Appendix A. ~$\hfill\blacksquare$
	
Based on Proposition \ref{pro:2D}, we can reexpress $q_3$, $q_6$, and $q_9$ as $\bar q_3 = \cos {\varphi _x}\sin {\varphi _y}$, $\bar q_6 = - \sin {\varphi _x}$, and $\bar q_9 = \cos {\varphi _x}\cos {\varphi _y}$, respectively. Let $\bar F \left(\varphi _x, \varphi _y\right) = {\cos {\theta _{\mathrm{I}}}}{{\cos }} {\theta _{\mathrm{T}}^{\mathrm{r}}} = (\mathbf{q}_z^T \bar{\mathbf{p}}_{\mathrm{I}})(\mathbf{q}_z^T \bar{\mathbf{p}}_{\mathrm{T}})$ with $\mathbf{q}_{z} = (\bar q_3,\bar q_6,\bar q_9)^T$, $\bar{\mathbf{p}}_{\mathrm{I}} = \frac{{\mathbf{p}}_{\mathrm{I}}}{\left\|{\mathbf{p}}_{\mathrm{I}}\right\|}$, and $\bar{\mathbf{p}}_{\mathrm{T}} = \frac{{\mathbf{p}}_{\mathrm{I}}-{\mathbf{p}}_{\mathrm{T}}}{\left\|{\mathbf{p}}_{\mathrm{I}}-{\mathbf{p}}_{\mathrm{T}}\right\|}$. 
Then, problem \eqref{pro:CRB} is reformulated as
\begin{subequations}
	\label{pro:CRB2}
	\begin{align}
		\mathop {\max }\limits_{\varphi_x, \varphi_y} \; &\bar F \left(\varphi _x, \varphi _y\right) \\
		\mathrm{s.t.} \; \;
		& {x_{\mathrm{I}}}{\bar q_3} + {y_{\mathrm{I}}}{\bar q_6} + {z_{\mathrm{I}}}{\bar q_9} \ge 0, \label{BS2D} \\
		& ( {{x_{\mathrm{I}}} \!-\! {x_{\mathrm{T}}}} ){\bar q_3} \!+\! ( {{y_{\mathrm{I}}} \!-\! {y_{\mathrm{T}}}} ){\bar q_6} \!+\! ( {{z_{\mathrm{I}}} \!-\! {z_{\mathrm{T}}}} ){\bar q_9} \ge 0. \label{target2D}
	\end{align}
\end{subequations}
	
Let $\mathbf{\Delta}_1 = (\mathbf{\Delta}_{1,x},\mathbf{\Delta}_{1,y},\mathbf{\Delta}_{1,z})^T = \bar{\mathbf{p}}_{\mathrm{I}}+ \bar{\mathbf{p}}_{\mathrm{T}}$, we provide the optimal solution to problem \eqref{pro:CRB2} in the following proposition.
\begin{Proposition}
	\label{pro:sensing}
	For sensing-only case, the optimal rotation angles are given by
	\begin{align}
		\varphi_x^\star = -\arcsin \frac{\mathbf{\Delta}_{1,y}}{\|\mathbf{\Delta}_{1}\|}, \;\;
		\varphi_y^\star = \arctan \frac{\mathbf{\Delta}_{1,x}}{\mathbf{\Delta}_{1,z}}. \label{phi_sen}
	\end{align}
\end{Proposition}
{\it{Proof:}} Please refer to Appendix B. ~$\hfill\blacksquare$
	
	
\subsection{Communication-Only Case}
Next, for the communication-only case, the SNR maximization problem is given by
\begin{subequations}
	\label{pro:communication}
	\begin{align}
		\mathop {\max}\limits_{\mathbf{\Phi}, \mathbf{W}_\mathrm{U},\bm{\varphi}} \;\; & {{{F_\mathrm{U}}\left( {\bm{\varphi }} \right)}{{{ {{\mathbf{h}}_\mathrm{U}^H{\mathbf{\Phi }}{\bar{\mathbf{H}}_{{\mathrm{BI}}}}{{\mathbf{W}}_\mathrm{U}}}\bar{\mathbf{H}}_{{\mathrm{BI}}}^H \mathbf{\Phi }^H {\mathbf{h}}_\mathrm{U}}}}} \\
		\mathrm{s.t.} \;\;\;\; 
		& \operatorname{tr}\left(\mathbf{W}_\mathrm{U}\right)  \le P_0,\\
		&\mathbf{W}_\mathrm{U} \succeq 0, \\
		& \operatorname{rank} \left(\mathbf{W}_\mathrm{U}\right) = 1, \\
		& ( {{x_{\mathrm{I}}} \!-\! {x_{\mathrm{U}}}} ){q_3} \!+\! ( {{y_{\mathrm{I}}} \!-\! {y_{\mathrm{U}}}} ){q_6} + ( {{z_{\mathrm{I}}} \!-\! {z_{\mathrm{U}}}} ){q_9} \ge 0, \\
		&\eqref{BSrotation}, \eqref{rank1}.
	\end{align}
\end{subequations}
	
Given $\mathbf{\Phi}$ and $\bm{\varphi}$, the optimal beamforming at the BS is given by
\begin{align}
	\label{BS_BF_C}
	{\mathbf{W}}_{\mathrm{U}}^{\star} = {P_0}\frac{{{{\mathbf{a}}^*_{{\mathrm{BI}}}}\left( {{\theta _{{\mathrm{BI}}}}} \right){\mathbf{a}}_{{\mathrm{BI}}}^T\left( {{\theta _{{\mathrm{BI}}}}} \right)}}{{\left\| {{{\mathbf{a}}_{{\mathrm{BI}}}}\left( {{\theta _{{\mathrm{BI}}}}} \right)} \right\|^2}}.
\end{align}

Given ${\mathbf{W}}_{\mathrm{U}}$ and $\bm{\varphi}$, the optimal SIRS phase shift is given by
\begin{align}
	\label{IRS_BF_C}
	\phi_n^{\star} = -\arg \left[{{\mathbf{a}}}_\mathrm{U}\left( \theta _{\mathrm{I,U}} \right) \right]_n  - \arg \left[ {{{\mathbf{a}}_{{\mathrm{IB}}}}\left( {{\vartheta _{{\mathrm{IB}}}}} \right)} \right]_n , \forall n \in \mathcal{N}.
\end{align}
	
Given $\mathbf{\Phi}^{\star}$ and ${\mathbf{W}}_{\mathrm{U}}^{\star}$, the objective function of problem \eqref{pro:communication} is reduced to $F_\mathrm{U}\left( {\bm{\varphi }} \right)$. Similar to Proposition \ref{pro:2D}, it can be readily verified that the SIRS's 2D orientation is sufficient to achieve the optimal performance of 3D orientation and the optimal rotation angles are given by
\begin{align}
	\varphi_x^\star = -\arcsin \frac{\mathbf{\Delta}_{2,y}}{\|\mathbf{\Delta}_{2}\|}, \;\;
	\varphi_y^\star = \arctan \frac{\mathbf{\Delta}_{2,x}}{\mathbf{\Delta}_{2,z}}, \label{phi_com}
\end{align}
where $\mathbf{\Delta}_2 = (\mathbf{\Delta}_{2,x},\mathbf{\Delta}_{2,y},\mathbf{\Delta}_{2,z})^T = \bar{\mathbf{p}}_{\mathrm{I}}+\bar{\mathbf{p}}_{\mathrm{U}}$ with $\bar{\mathbf{p}}_{\mathrm{U}} = \frac{{\mathbf{p}}_{\mathrm{I}}-{\mathbf{p}}_{\mathrm{U}}}{\left\|{\mathbf{p}}_{\mathrm{I}}-{\mathbf{p}}_{\mathrm{U}}\right\|}$.
	
\subsection{Co-located User and Target Case}
To draw more useful insight, we consider a special case where the sensed target also acts as a communication user. A typical example is a UAV or connected vehicle that receives downlink data from the BS while being sensed for localization and trajectory tracking \cite{colocated}. When the user and the target are co-located, their departure angles from the SIRS are the same, which yields $\mathbf a_{\mathrm U}(\theta_{\mathrm I,U}) = \mathbf a_{\mathrm{IT}}(\theta_{\mathrm{IT}})$ and $F_{\mathrm U}(\bm\varphi)
= F_{\mathrm T}(\bm\varphi)$. Based on \eqref{BS_BF_S}, \eqref{IRS_BF_S}, \eqref{BS_BF_C}, and \eqref{IRS_BF_C}, the SNR-constrained CRB minimization problem is formulated as
\begin{subequations}
	\label{pro:co}
	\begin{align}
		\mathop {\max }\limits_{\varphi_x, \varphi_y} \;\;\; &{\cos {\theta _{\mathrm{I}}}}{{\cos }} {\theta _{\mathrm{T}}^{\mathrm{r}}} \\
		\mathrm{s.t.} \;\;\;\;
		& {\cos {\theta _{\mathrm{I}}}}{{\cos }} {\theta _{\mathrm{T}}^{\mathrm{r}}}{|\alpha _{{\mathrm{U}}}|^2}{|\alpha _{{\mathrm{BI}}}|^2}P_0N^2M/{\sigma _{{\mathrm{U}}}^2} \ge \Gamma, \label{con:commu}\\
		& \eqref{BS2D}, \eqref{target2D}.
	\end{align}
\end{subequations}
	
Let $\bar \alpha$ denote the angle between $\mathbf{\bar{p}}_{\mathrm I}$ and $\mathbf{\bar{p}}_{\mathrm T}$, i.e., $\cos \bar{\alpha} = \bar{\mathbf{p}}_{\mathrm{I}}^T \bar{\mathbf{p}}_{\mathrm{T}} \in (-1,1)$. The closed-form expressions of the optimized solution to problem \eqref{pro:co} are derived in the following proposition.
\begin{Proposition}
	\label{proposition:co}
	If the condition $\cos \bar{\alpha} \ge 2\bar{\Gamma} - 1$
	is satisfied with $\bar \Gamma = {\sigma _{{\mathrm{U}}}^2}\Gamma / ({|\alpha _{{\mathrm{U}}}|^2}{|\alpha _{{\mathrm{BI}}}|^2}P_0N^2M)$, the optimal rotation angles are given by
	\begin{align}
		\varphi_x^\star = -\arcsin \frac{\mathbf{\Delta}_{1,y}}{\|\mathbf{\Delta}_1\|}, \;\;
		\varphi_y^\star = \arctan \frac{\mathbf{\Delta}_{1,x}}{\mathbf{\Delta}_{1,z}}.
	\end{align}
	\end{Proposition}
{\it{Proof:}} Please refer to Appendix C. ~$\hfill\blacksquare$
	
\subsection{Single User Case}
In this case, to independently characterize the impact of the spatial correlation on the ISAC performance, we consider a widely adopted idealized assumption with $F_{\mathrm{T}}(\bm{\varphi}) \approx 1$ and $F_{\mathrm{U}}(\bm{\varphi}) \approx 1$. Then, the optimization problem is formulated as 
\begin{subequations}
	\label{pro:STSU}
	\begin{align}
		\mathop {\max }\limits_{\substack{\mathbf{\Phi}, \mathbf{W}_\mathrm{U}, \\ \mathbf{R}_\mathrm{s}, \bm{\varphi}}}  &{{\left| {{\mathbf{a}}_{{\mathrm{IT}}}^T{\left( \bar \vartheta _{{\mathrm{TI}}} \right)}{\mathbf{\Phi }}{{\mathbf{a}}_{{\mathrm{IB}}}}\left( {{\vartheta _{{\mathrm{IB}}}}} \right)} \right|^2}{\mathbf{a}}_{{\mathrm{BI}}}^T\left( {{\theta _{{\mathrm{BI}}}}} \right){{\mathbf{R}}_{\mathrm{x}}}{{\mathbf{a}}^*_{{\mathrm{BI}}}}\left( {{\theta _{{\mathrm{BI}}}}} \right)}  \\
		\mathrm{s.t.} \;\; & {{{{{ {{\mathbf{h}}_\mathrm{U}^H{\mathbf{\Phi }}{\bar{\mathbf{H}}_{{\mathrm{BI}}}}{{\mathbf{W}}_\mathrm{U}}}\bar{\mathbf{H}}_{{\mathrm{BI}}}^H \mathbf{\Phi }^H {\mathbf{h}}_\mathrm{U}}}}}}\ge \sigma _\mathrm{U}^2 \Gamma_\mathrm{U}, \\
		& \operatorname{tr}\left(\mathbf{W}_\mathrm{U}\right) + \operatorname{tr}\left(\mathbf{R}_\mathrm{s}\right) \le P_0, \label{P0}\\
		& \mathbf{R}_\mathrm{s} \succeq 0, \label{Rs2}\\
		&\mathbf{W}_\mathrm{U} \succeq 0, \label{Wu}\\
		& \operatorname{rank} \left(\mathbf{W}_\mathrm{U}\right) = 1, \label{rankWu}\\
		& \eqref{BSrotation}, \eqref{targetrotation}, \eqref{userrotation}, \eqref{rank1}.
	\end{align}
\end{subequations}

Given $\mathbf{\Phi}$ and $\bm{\varphi}$, problem \eqref{pro:STSU} is reduced to
\begin{subequations}
	\label{pro:STSU_tr}
	\begin{align}
		\mathop {\max }\limits_{\mathbf{W}_\mathrm{U},\mathbf{R}_\mathrm{s}} \; & {\mathbf{a}}_{{\mathrm{BI}}}^T\left( {{\theta _{{\mathrm{BI}}}}} \right){{\mathbf{R}}_{\mathrm{x}}}{{\mathbf{a}}^*_{{\mathrm{BI}}}}\left( {{\theta _{{\mathrm{BI}}}}} \right)  \\
		\mathrm{s.t.} \;\;\; & {\left| {{\tilde {\alpha} _{{0}}}} \right|^2}{\left| {{\alpha _{\mathrm{U}}}} \right|^2}{\left| {{\mathbf{a}}_{\mathrm{U}}^T\left( {{\theta _{{\mathrm{I}},{\mathrm{U}}}}} \right){\mathbf{\Phi }}{{\mathbf{a}}_{{\mathrm{IB}}}}\left( {{\vartheta _{{\mathrm{IB}}}}} \right)} \right|^2} \nonumber\\
		&\times {\mathbf{a}}_{{\mathrm{BI}}}^T\left( {{\theta _{{\mathrm{BI}}}}} \right){{\mathbf{W}}_{\mathrm{U}}}{{\mathbf{a}}^*_{{\mathrm{BI}}}}\left( {{\theta _{{\mathrm{BI}}}}} \right) \ge \sigma _\mathrm{U}^2{\Gamma _{\mathrm{U}}}, \\
		& \eqref{P0}, \eqref{Rs2}, \eqref{Wu}, \eqref{rankWu}.
	\end{align}
\end{subequations}

According to \cite{songISAC}, the optimal solution to the relaxed SDR version of problem \eqref{pro:STSU_tr} is given by
\begin{align}
	\mathbf{W}_\mathrm{U}^\star &= \bar{\mathbf{w}}_\mathrm{U} \bar{\mathbf{w}}_\mathrm{U}^H, \forall k \in \mathcal{K}, \\
	\mathbf{R}_\mathrm{s}^\star &= \bar{\mathbf{R}}_\mathrm{s} +  \bar{\mathbf{W}}_\mathrm{U} -  \mathbf{W}_\mathrm{U}^\star,
\end{align}
where $\bar{\mathbf{w}}_\mathrm{U} = ({\mathbf{a}}_{{\mathrm{BI}}}^T \bar{\mathbf{W}}_\mathrm{U} {\mathbf{a}}^*_{{\mathrm{BI}}} )^{-1/2} \bar{\mathbf{W}}_\mathrm{U} {\mathbf{a}}^*_{{\mathrm{BI}}}$. 
There exists an optimal solution $\bar{\mathbf{R}}_\mathrm{s}^\star = 0$. Since the rank bound of an optimal solution $\mathbf{R}_\mathrm{x}^\star$ to problem \eqref{pro:STSU_tr} is given by $\operatorname{rank}^2 \left(\mathbf{R}_\mathrm{x}^\star\right) \le 2$ \cite{beam}, we have $\operatorname{rank} \left(\mathbf{R}_\mathrm{x}^\star\right) = 1$. Then, the optimal solutions to problem \eqref{pro:STSU_tr} are given by
\begin{align}
	\mathbf{R}_\mathrm{s}^\star = 0, \;\;
	\mathbf{w}_\mathrm{U}^\star = \sqrt{P_0} \frac{{\mathbf{a}}^*_{{\mathrm{BI}}}\left( {{\theta _{{\mathrm{BI}}}}} \right)}{\|{\mathbf{a}}_{{\mathrm{BI}}}\left( {{\theta _{{\mathrm{BI}}}}} \right)\|}.
\end{align}
Under the considered setup, the communication signal can illuminate the target and provide the sensing measurements. Consequently, the transmit power required for sensing can be carried by the communication beam, and no dedicated sensing signal is needed. Allocating all transmit power to the communication component does not sacrifice the sensing capability captured by the model.

Given $\mathbf{w}_\mathrm{U}^\star$ and $\mathbf{R}_\mathrm{x}^\star$, problem \eqref{pro:STSU} is reduced to
\begin{subequations}
	\label{pro:STSU1}
	\begin{align}
		\mathop {\max }\limits_{\bm{\varphi}, \mathbf{\Phi}} \;\; &  {\left| {{\mathbf{a}}_{{\mathrm{IT}}}^T{\left( \bar \vartheta _{{\mathrm{TI}}} \right)}{\mathbf{\Phi }}{{\mathbf{a}}_{{\mathrm{IB}}}}\left( {{\vartheta _{{\mathrm{IB}}}}} \right)} \right|^2}  \\
		\mathrm{s.t.} \;\; 
		& {\left| {{\mathbf{a}}_{\mathrm{U}}^T\left( {{\theta _{{\mathrm{I}},{\mathrm{U}}}}} \right){\mathbf{\Phi }}{{\mathbf{a}}_{{\mathrm{IB}}}}\left( {{\vartheta _{{\mathrm{IB}}}}} \right)} \right|^2} \ge {\bar{\Gamma} _{\mathrm{U}}}, \label{con:snr_c} \\
		& \eqref{BSrotation}, \eqref{targetrotation}, \eqref{userrotation}, \eqref{rank1},
	\end{align}
\end{subequations}
where ${\bar{\Gamma} _{\mathrm{U}}} = \frac{\sigma _\mathrm{U}^2{\Gamma _{\mathrm{U}}}}{P_0 M {\left| {{\tilde {\alpha} _{{0}}}} \right|^2}{\left| {{\alpha _{\mathrm{U}}}} \right|^2}}$.

With the optimal SIRS beamforming vector, the objective function of problem \eqref{pro:STSU1} is given by \cite{RA_UAV}
\begin{align}
	g_1 \!=\! \left\{\!\!\! {\begin{array}{*{20}{c}}
			N^2, &{{\rho ^2}{N^2} \ge {{\bar \Gamma }_{\mathrm{U}}},}\\
				\left(\rho \sqrt{\frac{\bar{\Gamma}_{\mathrm{U}}}{\|\mathbf{u}_{\mathrm{U}}\|^2}} \!+\! \sqrt{N^2 \!-\! \frac{\bar{\Gamma}_{\mathrm{U}}}{\|\mathbf{u}_{\mathrm{U}}\|^2}} \sqrt{1 \!-\! \rho^2}\right)^2,&{{\mathrm{otherwise,}}}
	\end{array}} \right.
    \label{lemma}
\end{align}
where $\rho = \frac{|\mathbf{u}_{\mathrm{U}}^H \mathbf{u}_{\mathrm{T}}|}{\|\mathbf{u}_{\mathrm{U}}\| \|\mathbf{u}_{\mathrm{T}}\|} \in [0, 1]$ denotes the correlation coefficient, with $\mathbf{u}_{\mathrm{T}} \triangleq \operatorname{diag}\left(\mathbf{a}_{\mathrm{IT}}^T(\vartheta_{\mathrm{TI}})\right) \mathbf{a}_{\mathrm{IB}}(\vartheta_{\mathrm{IB}})$, and $\mathbf{u}_{\mathrm{U}} \triangleq \operatorname{diag}\left(\mathbf{a}_{\mathrm{U}}^T(\theta_{\mathrm{I,U}})\right) \mathbf{a}_{\mathrm{IB}}(\vartheta_{\mathrm{IB}})$. It indicates that the rotation angle should be designed to maximize $\rho$ for enhancing the sensing and communication performance through channel subspace rotation. 
In the following proposition, we provide the optimal solution to problem \eqref{pro:STSU1}.
\begin{Proposition} 
	\label{proposition:subspace}
	The optimal SIRS phase shift is given by 
	\begin{align}
		\phi _n^* = -{\arg {{\left[ {{{\bf{a}}_{{\mathrm{IT}}}}\left( {{\vartheta _{{\mathrm{TI}}}}} \right)} \right]}_n} \!-\! \arg {{\left[ {{{\bf{a}}_{{\mathrm{IB}}}}\left( {{\vartheta _{{\mathrm{IB}}}}} \right)} \right]}_n}}, \forall n \in \mathcal{N}.
	\end{align}
	The optimal rotation angles are given by
	\begin{align}
		\varphi_x^\star = -\arcsin \frac{\mathbf{\Delta}_{3,y}}{\|\mathbf{\Delta}_3\|}, \;\;
		\varphi_y^\star = \arctan \frac{\mathbf{\Delta}_{3,x}}{\mathbf{\Delta}_{3,z}},
	\end{align}
	where $\mathbf{\Delta}_3 = (\mathbf{\Delta}_{3,x},\mathbf{\Delta}_{3,y},\mathbf{\Delta}_{3,z})^T$ and $\mathbf{q}_z^* = \frac{\mathbf{\Delta}_{3}}{\|\mathbf{\Delta}_3\|}$ within the feasible set $\mathcal{Q} = \{ \mathbf{q}_z | \mathbf{q}_z^T \mathbf{c}_\mathrm{UT} = 0, \|\mathbf{q}_z\| = 1, \mathbf{q}_z^T \bar{\mathbf{p}}_\mathrm{I} \ge 0, \mathbf{q}^T \bar{\mathbf{p}}_\mathrm{T} \ge 0, \mathbf{q}_z^T \bar{\mathbf{p}}_\mathrm{U} \ge 0 \}$ with $\mathbf{c}_\mathrm{UT} = \bar{\mathbf{p}}_{\mathrm{U}} - \bar{\mathbf{p}}_{\mathrm{T}}$. 
\end{Proposition}
{\it{Proof:}} Please refer to Appendix D. ~$\hfill\blacksquare$
	
In the following, we restore the practical orientation-dependent aperture gains and consider the general rotation design. Given $\mathbf{W}_\mathrm{U}$, $\mathbf{R}_\mathrm{s}$, and $\mathbf{\Phi}$, problem \eqref{pro:original} is reduced to 
\begin{subequations}
	\label{pro:q}
	\begin{align}
		\mathop {\max }\limits_{ \mathbf{q}_z } \; &(\mathbf{q}_z^T \bar{\mathbf{p}}_{\mathrm{I}}) (\mathbf{q}_z^T \bar{\mathbf{p}}_{\mathrm{T}})  {\left| {{\mathbf{a}}_{{\mathrm{IT}}}^T{\left( \bar \vartheta _{{\mathrm{TI}}} \right)}{\mathbf{\Phi }}{{\mathbf{a}}_{{\mathrm{IB}}}}\left( {{\vartheta _{{\mathrm{IB}}}}} \right)} \right|^2}  \\
		\mathrm{s.t.} \;\; &  \frac{1}{\Gamma_\mathrm{U}}{ (\mathbf{q}_z^T \bar{\mathbf{p}}_{\mathrm{I}}) (\mathbf{q}_z^T \bar{\mathbf{p}}_\mathrm{U})  }|\alpha_\mathrm{U}|^2 |\alpha_{\mathrm{BI}}|^2 |\mathbf{a}_\mathrm{U}^T(\theta_{\mathrm{I},\mathrm{U}}) \mathbf{\Phi} \mathbf{a}_{\mathrm{IB}}(\vartheta_{\mathrm{IB}})|^2 \nonumber\\ 
		&\times \mathbf{a}_{\mathrm{BI}}^T(\theta_{\mathrm{BI}}) \mathbf{W}_\mathrm{U} \mathbf{a}^*_{\mathrm{BI}}(\theta_{\mathrm{BI}}) 
		- \sigma _\mathrm{U}^2 \ge 0, \label{con:snr} \\
		& \mathbf{q}_z^T {\mathbf{p}}_{\mathrm{I}} \ge 0, \label{qi}\\
		& \mathbf{q}_z^T ({\mathbf{p}}_{\mathrm{I}} - {\mathbf{p}}_{\mathrm{T}}) \ge 0, \label{qit}\\
		& \mathbf{q}_z^T ({\mathbf{p}}_{\mathrm{I}} - {\mathbf{p}}_\mathrm{U}) \ge 0, \label{qik}\\
		& \|\mathbf{q}_z\| = 1. \label{con:q}
	\end{align}
\end{subequations}
Let $b_\mathrm{T} (\mathbf{q}_z) = {\mathbf{a}}_{{\mathrm{IT}}}^T{ ( \bar \vartheta _{{\mathrm{TI}}} )} {\mathbf{\Phi }}{{\mathbf{a}}_{{\mathrm{IB}}}} ( {{\vartheta _{{\mathrm{IB}}}}}  )$ and $b_\mathrm{U} (\mathbf{q}_z) = \mathbf{a}_\mathrm{U}^T(\theta_{\mathrm{I},\mathrm{U}}) \mathbf{\Phi} \\ \times \! \mathbf{a}_{\mathrm{IB}}(\vartheta_{\mathrm{IB}})$. Then, we have ${{\dot b}_{\mathrm{T}}}({{\bf{q}}_z}) \!=\! \frac{{\partial |{b_{\mathrm{T}}}({{\bf{q}}_z}){|^2}}}{{\partial {{\bf{q}}_z}}} \!=\! 2 \operatorname{Re}\{ b_{\mathrm{T}}^*({{\bf{q}}_z}) \\ \times (j\frac{{\pi {d_\mathrm{I}}}}{\lambda }\cos {\vartheta _{{\mathrm{IB}}}}({\bf{a}}_{{\mathrm{IT}}}^T({{\bar \vartheta }_{{\mathrm{TI}}}}){\bf{\Phi D}}{{\bf{a}}_{{\mathrm{IB}}}}({\vartheta _{{\mathrm{IB}}}})){{{\bf{\tilde p}}}_{{\mathrm{IB}}}} + j\frac{{\pi {d_\mathrm{I}}}}{\lambda } \cos {\theta _{{\mathrm{IT}}}} \\ \times ({\bf{a}}_{{\mathrm{IT}}}^T({{\bar \vartheta }_{{\mathrm{TI}}}}){\bf{D\Phi }}{{\bf{a}}_{{\mathrm{IB}}}}({\vartheta _{{\mathrm{IB}}}})){{{\bf{\tilde p}}}_{{\mathrm{IT}}}})\}$ and ${{\dot b}_\mathrm{U}}({{\bf{q}}_z}) = \frac{{\partial |{b_{\mathrm{U}}}({{\bf{q}}_z}){|^2}}}{{\partial {{\bf{q}}_z}}} = 2\operatorname{Re}\{ b_\mathrm{U}^*({{\bf{q}}_z}) ( j\frac{{\pi{d_{\mathrm{I}}}}}{\lambda }\cos {\vartheta _{{\mathrm{IB}}}}({\bf{a}}_\mathrm{U}^T(\theta_{\mathrm{I},\mathrm{U}}) {\bf{\Phi D}} {{\bf{a}}_{{\mathrm{IB}}}}(\vartheta_{\mathrm{IB}})){{{\bf{\tilde p}}}_{{\mathrm{IB}}}} + j\frac{{\pi{d_{\mathrm{I}}}}}{\lambda } \\ \times \cos {\theta _{{\mathrm{I}},\mathrm{U}}}({\bf{a}}_\mathrm{U}^T(\theta_{\mathrm{I},\mathrm{U}}){\bf{D\Phi }}{{\bf{a}}_{{\mathrm{IB}}}}(\vartheta_{\mathrm{IB}})){{{\bf{\tilde p}}}_\mathrm{U}} ) \}$ with ${{{\bf{\tilde p}}}_{{\mathrm{IB}}}} = \frac{{ - {{\bf{p}}_{\mathrm{I}}}}}{{\sqrt {{{(x_{\mathrm{B}}^{\mathrm{L}})}^2} + {{(y_{\mathrm{B}}^{\mathrm{L}})}^2}} }}$, ${{{\bf{\tilde p}}}_{{\mathrm{IT}}}} =  - \frac{{{{\bf{p}}_{\mathrm{T}}} - {{\bf{p}}_{\mathrm{I}}}}}{{\sqrt {{{(x_{\mathrm{T}}^{\mathrm{L}})}^2} + {{(y_{\mathrm{T}}^{\mathrm{L}})}^2}} }}$, ${{{\bf{\tilde p}}}_\mathrm{U}} =  - \frac{{{{\bf{p}}_\mathrm{U}} - {{\bf{p}}_{\mathrm{I}}}}}{{\sqrt {{{(x_\mathrm{U}^{\mathrm{L}})}^2} + {{(y_\mathrm{U}^{\mathrm{L}})}^2}} }}$, and ${\bf{D}} = \operatorname{diag} ( - (N - 1), - (N - 3), \cdots ,N - 1)$. The objective function of problem \eqref{pro:q} is rewritten as $f_1 (\mathbf{q}_z) =  {\left| b_\mathrm{T} (\mathbf{q}_z) \right|^2} \\ \times (\mathbf{q}_z^T \bar{\mathbf{p}}_{\mathrm{I}}) (\mathbf{q}_z^T \bar{\mathbf{p}}_{\mathrm{T}}) $. Constraint \eqref{con:snr} is expressed as $f_{2} (\mathbf{q}_z) - \sigma _\mathrm{U}^2 \ge 0$, where $f_{2} (\mathbf{q}_z) =  \frac{B_\mathrm{U}}{\Gamma_\mathrm{U}} {(\mathbf{q}_z^T \bar{\mathbf{p}}_{\mathrm{I}}) (\mathbf{q}_z^T \bar{\mathbf{p}}_\mathrm{U}) } {{ |b_\mathrm{U}\left(\mathbf{q}_z\right)|^2 }}$ with $B_\mathrm{U} = |\alpha_\mathrm{U}|^2 |\alpha_{\mathrm{BI}}|^2 \mathbf{a}_{\mathrm{BI}}^T(\theta_{\mathrm{BI}}) \mathbf{W}_\mathrm{U} \mathbf{a}^*_{\mathrm{BI}}(\theta_{\mathrm{BI}})$. By using the first-order Taylor approximation, for the given local point $\mathbf{q}_z^{(r)}$ in the $r$-th iteration, we have 
\begin{align}
	f_1 (\mathbf{q}_z) 
	&\approx f_1 (\mathbf{q}_z^{(r)}) + \left(  |b_{\mathrm{T}}(\mathbf{q}_z^{(r)})|^2  \bar{\mathbf{p}}_{\mathrm{I}} ((\mathbf{q}_z^{(r)})^T \bar{\mathbf{p}}_{\mathrm{T}}) \right. \nonumber\\
	&+ |b_{\mathrm{T}}(\mathbf{q}_z^{(r)})|^2 \bar{\mathbf{p}}_{\mathrm{T}} ((\mathbf{q}_z^{(r)})^T \bar{\mathbf{p}}_{\mathrm{I}}) \nonumber\\
	&\left. + ((\mathbf{q}_z^{(r)})^T \bar{\mathbf{p}}_{\mathrm{I}}) ((\mathbf{q}_z^{(r)})^T \bar{\mathbf{p}}_{\mathrm{T}}) \dot b_\mathrm{T} (\mathbf{q}_z^{(r)}) \right)^T (\mathbf{q}_z - \mathbf{q}_z^{(r)}) \nonumber \\
	&\triangleq \hat f_1^{(r)} (\mathbf{q}_z), \\
	f_{2} (\mathbf{q}_z)
	&\approx f_{2} (\mathbf{q}_z^{(r)}) + \left(  \frac{B_\mathrm{U}}{\Gamma_\mathrm{U}}  \right. ( |b_\mathrm{U}(\mathbf{q}_z^{(r)})|^2 ( \bar{\mathbf{p}}_{\mathrm{I}} ( (\mathbf{q}_z^{(r)})^T \bar{\mathbf{p}}_\mathrm{U}) \nonumber\\
	&+ ( (\mathbf{q}_z^{(r)})^T \bar{\mathbf{p}}_{\mathrm{I}}) \bar{\mathbf{p}}_\mathrm{U} ) 
	+ ((\mathbf{q}_z^{(r)})^T \bar{\mathbf{p}}_{\mathrm{I}}) ((\mathbf{q}_z^{(r)})^T \bar{\mathbf{p}}_\mathrm{U}) \nonumber\\
	&\times \left.  \dot b_\mathrm{U} (\mathbf{q}_z^{(r)}) ) \right)^T (\mathbf{q}_z - \mathbf{q}_z^{(r)}) \nonumber\\
	&\triangleq \hat f_{2}^{(r)} (\mathbf{q}_z).
\end{align}

By relaxing the unit-modulus constraint in \eqref{con:q}, problem \eqref{pro:q} can be approximated by
\begin{subequations}
	\label{pro:q1}
	\begin{align}
		\mathop {\max }\limits_{ \mathbf{q}_z } \;\; &\hat f_1^{(r)} (\mathbf{q}_z)  \\
		\mathrm{s.t.} \;\;\;
		&  \hat f_{2}^{(r)} (\mathbf{q}_z) - \sigma _\mathrm{U}^2 \ge 0, \\
		& \eqref{qi}, \eqref{qit}, \eqref{qik},
	\end{align}
\end{subequations}
which is a convex optimization problem and thus can be solved by CVX. Let $\hat{\mathbf{q}}_z$ denote the optimal solution to problem \eqref{pro:q1}. If $\hat{\mathbf{q}}_z$ does not satisfy $\| \hat{\mathbf{q}}_z \| = 1$, we reconstruct a high-quality solution to problem \eqref{pro:q}, i.e., $\mathbf{q}_z^\star = \frac{\hat{\mathbf{q}}_z}{\| \hat{\mathbf{q}}_z \|} = ( q_3^*, q_6^*, q_9^*)^T$.

\subsection{Multi-User Case}
For the general case with the multi-path BS-SIRS channel model and multiple users, we divide all the variables into three blocks, i.e., SIRS rotation $\bm{\varphi}$, BS beamformer $\mathbf{W}_{k}$ and $\mathbf{R}_\mathrm{s}$, as well as SIRS phase-shifts, and then each block is optimized iteratively until convergence is achieved.
\subsubsection{Rotation Optimization}
Given $\mathbf{W}_{k}$, $\mathbf{R}_\mathrm{s}$, and $\mathbf{\Phi}$, problem \eqref{pro:original} is reduced to
\begin{align}
		\label{pro:STMU_rotate}
		\mathop {\min }\limits_{\bm{\varphi}} \operatorname{CRB}{\left( \bar \vartheta _{{\mathrm{TI}}} \right)} \;\;\;
		\mathrm{s.t.} \; \eqref{sinr},\eqref{BSrotation},\eqref{targetrotation},\eqref{userrotation}.
\end{align}
The optimal solution to problem \eqref{pro:STMU_rotate} can be obtained through a two-dimensional search over $\varphi_x \in [-90^\circ,90^\circ]$ and $\varphi_y \in [-90^\circ,90^\circ]$. Moreover, the search space can be reduced by considering the reflection constraints. To further reduce the computational complexity, a genetic algorithm can be employed. Specifically, each chromosome is represented by $\bm{\phi}=[\varphi_x,\varphi_y]^T$. The rotation angles that satisfy the reflection constraints and user SINR requirements are retained in the population. In each generation, the solutions with larger sensing information are selected as parents for the next generation using tournament selection. Blend crossover combines the two angular components of the selected parents, while Gaussian mutation perturbs each angle to preserve population diversity.

\subsubsection{BS Beamforming Optimization}
Let $\mathbf{G}_{k} = \mathbf{g}_k \mathbf{g}_k^H$ with $\mathbf{g}_k^H = {\mathbf{h}}_{k}^H{\mathbf{\Phi }}{{\mathbf{H}}_{{\mathrm{BI}}}}$. Given $\mathbf{\Phi}$ and $\bm{\varphi}$, problem \eqref{pro:original} is reduced to
\begin{subequations}
	\label{pro:TxBF}
	\begin{align}
		\mathop {\max }\limits_{ \left\{\mathbf{W}_{k}\right\},\mathbf{R}_\mathrm{s}} \; &{\operatorname{tr} \left( {{\mathbf{\dot{A}}}{{\mathbf{R}}_{\mathrm{x}}}{{{\mathbf{\dot{A}}}}^H}} \right) - \frac{{{{\left| {\operatorname{tr} \left( {{\mathbf{A}}{{\mathbf{R}}_{\mathrm{x}}}{{{\mathbf{\dot{A}}}}^H}} \right)} \right|}^2}}}{{\operatorname{tr} \left( {{\mathbf{A}}{{\mathbf{R}}_{\mathrm{x}}}{{\mathbf{A}}^H}} \right)}}}  \\
		\mathrm{s.t.} \;\;\; & \sigma _k^2 - \frac{1}{\Gamma_k}{{F_k}\left( {\bm{\varphi }} \right)} \operatorname{tr} \left(\mathbf{G}_{k}\mathbf{W}_{k}\right) \nonumber\\
		&+ \sum _{i \in \mathcal{K},i \ne k} {{F_k}\left( {\bm{\varphi }} \right)}\operatorname{tr} \left(\mathbf{G}_{k}\mathbf{W}_{i}\right) \le 0, \forall k \in \mathcal{K}, \label{con:BF_snr}\\
		&\eqref{txpower}, \eqref{Rs}, \eqref{wk}, \eqref{rankwk}.
	\end{align}
\end{subequations}
By introducing an auxiliary variable $g_2$ and using the Schur's complement, problem \eqref{pro:TxBF} is reformulated as
\begin{subequations}
	\label{pro:TxBF1}
	\begin{align}
		\mathop {\max }\limits_{ \left\{\mathbf{W}_{k}\right\},\mathbf{R}_\mathrm{s},g_2}  &g_2  \\
		\mathrm{s.t.} \;\;\;\;\;
		& \left[\!\! {\begin{array}{*{20}{c}}
				{\operatorname{tr} ( {{\mathbf{\dot A}}{{\mathbf{R}}_{\mathrm{x}}}{{{\mathbf{\dot A}}}^H}} ) - {g_2}}&{\operatorname{tr} ( {{\mathbf{A}}{{\mathbf{R}}_{\mathrm{x}}}{{{\mathbf{\dot A}}}^H}} )}\\
				{\operatorname{tr} ( {{\mathbf{\dot A}}{{\mathbf{R}}_{\mathrm{x}}}{{\mathbf{A}}^H}} )}&{\operatorname{tr} ( {{\mathbf{A}}{{\mathbf{R}}_{\mathrm{x}}}{{\mathbf{A}}^H}} )}
		\end{array}} \!\! \right] \succeq 0, \\
		& \eqref{con:BF_snr},\eqref{txpower}, \eqref{Rs}, \eqref{wk}, \eqref{rankwk}.
	\end{align}
\end{subequations}

By dropping the rank-one constraint, we obtain the relaxed version of problem \eqref{pro:TxBF1} as problem (SDR67), which is a convex semi-definite program (SDP) that can be optimally solved by convex solvers such as CVX. Let $\left\{\bar{\mathbf{W}}_{k}\right\}$ and $\bar{\mathbf{R}}_\mathrm{s}$ denote the optimal solution to (SDR67). According to \cite{songISAC}, the optimal solution to problem \eqref{pro:TxBF} is given by
\begin{align}
	\mathbf{W}_{k}^\star &= \bar{\mathbf{w}}_{k} \bar{\mathbf{w}}_{k}^H, \forall k \in \mathcal{K}, \label{W_k} \\
	\mathbf{R}_\mathrm{s}^\star &= \bar{\mathbf{R}}_\mathrm{s} + \sum_{k \in \mathcal{K}} \bar{\mathbf{W}}_{k} - \sum_{k \in \mathcal{K}} \mathbf{W}_{k}^\star, \label{R_s}
\end{align}
where $\bar{\mathbf{w}}_{k} = (\mathbf{g}_k^H \bar{\mathbf{W}}_{k} \mathbf{g}_k )^{-1/2} \bar{\mathbf{W}}_{k} \mathbf{g}_k$.

\subsubsection{IRS Beamforming Optimization}
Let ${\mathbf{h}}_k^H \mathbf{\Phi} {\mathbf{H}}_\mathrm{BI} = \mathbf{v}^T {\mathbf{H}}_k$ with ${\mathbf{H}}_k = \operatorname{diag}({\mathbf{h}}_k^H) {\mathbf{H}}_\mathrm{BI}$ and $\mathbf{V} = \mathbf{v} \mathbf{v}^H$ satisfying $\mathbf{V} \succeq 0$ and $\operatorname{rank} \left(\mathbf{V}\right) = 1$. Given $\mathbf{W}_{k}$, $\mathbf{R}_\mathrm{s}$, and $\bm{\varphi}$, problem \eqref{pro:original} is reduced to
\begin{subequations}
	\label{pro:IRSBF}
	\begin{align}
		\mathop {\max }\limits_{\mathbf{V}} \;\; & {\frac{{{L^2} \!-\! 1}}{3}} \! \operatorname{tr} ( {\mathbf{R}}_1 {\mathbf{V}} ) \!+\! \operatorname{tr} ( {{{\mathbf{D}}_{\mathrm{IT}}}{{\mathbf{R}}_1}{{\mathbf{D}}_{\mathrm{IT}}}{\mathbf{V}}} ) \!-\! \frac{{{\left| \operatorname{tr} ({{\mathbf{D}_\mathrm{IT}}{{\mathbf{R}}_1}{\mathbf{V}}}) \right|^2}}}{{\operatorname{tr}( {{{\mathbf{R}}_1}{\mathbf{V}}} )}} \\
		\mathrm{s.t.} \;\;
		&\sum _{i \in \mathcal{K},i \ne k} {{F_k}\left( {\bm{\varphi }} \right)}\operatorname{tr} \left(\mathbf{\hat H}_{k,i} \mathbf{V}\right) \nonumber\\
		&- \frac{1}{\Gamma_k}{{F_k}\left( {\bm{\varphi }} \right)} \operatorname{tr} \left(\mathbf{\hat H}_{k,k} \mathbf{V}\right) + \sigma _k^2 \le 0, \forall k \in \mathcal{K}, \label{V_snr}\\
		& \mathbf{V} \succeq 0, \label{V}\\
		&{\mathbf{V}}_{n,n}=1, \forall n \in \mathcal{N} \label{rankV},
	\end{align}
\end{subequations}
where $\mathbf{\hat H}_{k,i} = (\mathbf H_k\mathbf W_i\mathbf H_k^H)^T$.

By introducing an auxiliary variable $g_3$ and using the Schur's complement, problem \eqref{pro:IRSBF} is reformulated as
\begin{subequations}
	\label{pro:V}
	\begin{align}
		\mathop {\max }\limits_{\mathbf{V}} \;\; &( {\frac{{{L^2} - 1}}{3}} ) \operatorname{tr} ( {\mathbf{R}}_1 {\mathbf{V}} ) + \operatorname{tr} ( {{{\mathbf{D}}_{\mathrm{IT}}}{{\mathbf{R}}_1}{{\mathbf{D}}_{\mathrm{IT}}}{\mathbf{V}}} ) - g_3 \\
		\mathrm{s.t.} \;\; 
				& \left[ {\begin{array}{*{20}{c}}
				{{g_3}}&{\operatorname{tr} \left( {{{\mathbf{D}}_{{\mathrm{IT}}}}{{\mathbf{R}}_1}{\mathbf{V}}} \right)}\\
				{\operatorname{tr} \left( {{{\mathbf{V}}^H}{{\mathbf{R}}_1}{{\mathbf{D}}_{{\mathrm{IT}}}}} \right)}&{\operatorname{tr} \left( {{{\mathbf{R}}_1}{\mathbf{V}}} \right)}
		\end{array}} \right] \succeq \mathbf{0},\\
		&\eqref{V_snr}, \eqref{V}, \eqref{rankV}.
	\end{align}
\end{subequations}

By dropping the rank-one constraint, we obtain the relaxed version of problem \eqref{pro:V} as problem (SDR71), which is a convex SDP that can be optimally solved by convex solvers such as CVX. 

Overall Algorithm and Computational Complexity: In summary, the proposed algorithm updates the SIRS rotation angles $\bm{\varphi}$, the BS beamforming $\{\mathbf{W}_k\}$ and $\mathbf{R}_s$, and the SIRS beamforming $\bm{\Phi}$ alternately in an iterative manner. Specifically, $\bm{\varphi}$ is updated by applying the genetic algorithm to problem \eqref{pro:STMU_rotate}, while $\{\mathbf{W}_k\}$ and $\mathbf{R}_s$ are updated by solving problem (SDR67), followed by reconstruction in \eqref{W_k} and \eqref{R_s}. With the updated rotation angles and BS beamforming, the SIRS beamforming is optimized by solving problem (SDR71). Let $\mathbf{V}^\star$ denote the optimal solution to problem (SDR71). If $\mathbf{V}^\star$ is not a rank-one matrix, we first generate a number of random vectors $\mathbf{r} \sim \mathcal{CN} (0,\mathbf{I}_N)$and, for each random vector, construct a candidate solution $\mathbf {\tilde v} = e^{j \arg \{(\mathbf{V}^\star)^{1/2} \mathbf{r}\}}$, from which the reflection matrix is reconstructed as $\bm{\Phi}=\operatorname{diag}(\mathbf {\tilde v})$. The above procedures are repeated until the convergence is achieved. 

The complexities of problem (SDR67) and problem (SDR71) are $\mathcal O((KM)^{3.5})$ and $\mathcal O(N^{3.5})$, respectively. For rotation optimization, the complexity of the genetic algorithm is $\mathcal O(P_{\mathrm{GA}}G_{\mathrm{GA}} C_{\mathrm{GA}})$, where $P_{\mathrm{GA}}$ denotes the population size, $G_{\mathrm{GA}}$ denotes the number of generations, and $\mathcal C_{\mathrm{GA}}$ denotes the cost of evaluating one candidate. Thus, the main complexity of the proposed algorithm is given by $\mathcal{O} \left( L_{\mathrm{iter}} \left( P_{\mathrm{GA}}G_{\mathrm{GA}} C_{\mathrm{GA}} \!+\! (K M)^{3.5} \!+\! N^{3.5} \right)\right)$, where $L_{\mathrm{iter}}$ denotes the total number of iterations required to achieve convergence.

\section{Numerical Results}
In this section, we provide the simulations to demonstrate the performance of the rotatable SIRS-aided ISAC system. The global coordinates of the BS, SIRS, and target are $[0,0,0]^T$ m, $[20,30,15]^T$ m, and $[0,20,10]^T$ m, respectively. The users are uniformly and randomly distributed within a radius of 5 m centered at $[25,35,0]^T$ m. Without loss of generality, other
system parameters are set as follows: $M=128$, $N=256$, $\lambda = 0.125$ m, $\kappa = 7$ dBsm, $L=4$, $T=100$, $P_0 = 40$ dBm, $d_\mathrm{s} = d_\mathrm{I} = \bar d = \lambda /2$, $\Gamma_{\mathrm{U}} = \Gamma _k = 10$ dB, $\sigma_{\mathrm{U}} = \sigma _k = -80$ dBm, $\forall k \in \mathcal{K}$, $\sigma_\mathrm{T} = -110$ dBm, and $C=5$.

\subsection{Sensing-Only Case}
\begin{figure}[t]
	\centering
	\includegraphics[width=0.85\linewidth]{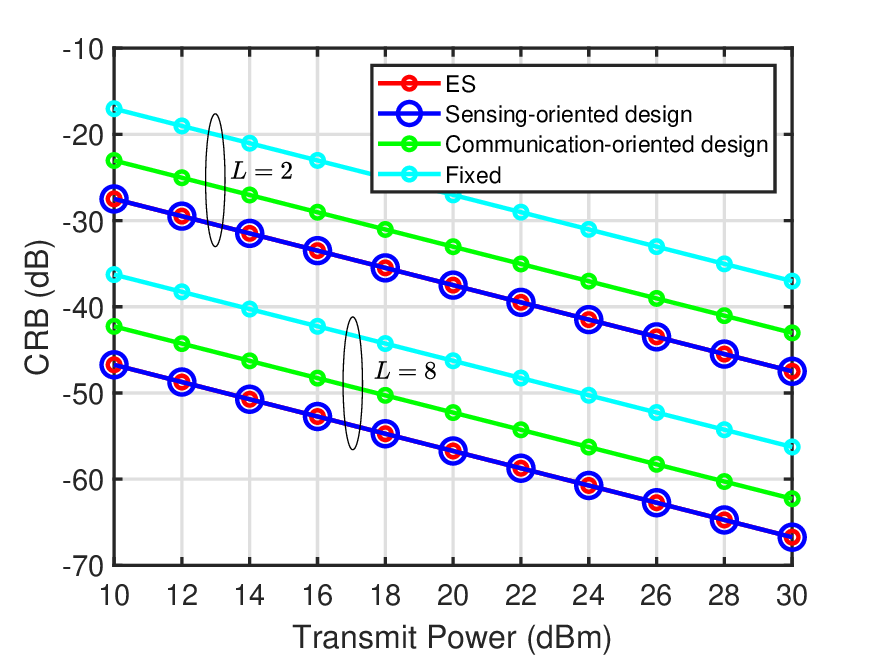}
	\vspace{-5pt}
	\caption{CRB versus the transmit power $P_0$.}
	\label{fig:CRB_P}
    \vspace{-5pt}
\end{figure}

We first investigate a special case of the ISAC system with a point target. For comparison, the following rotation design schemes are considered: 1) Sensing-oriented design: the SIRS rotation angles are set based on \eqref{phi_sen}; 2) Communication-oriented design: the SIRS rotation angles are set based on \eqref{phi_com}; 3) Fixed: the SIRS rotation angles are set to $\bm{\varphi} = \mathbf{0}$; 4) ES: the optimized SIRS rotation angles are obtained via the two-dimensional search.

In Fig. \ref{fig:CRB_P}, we plot the CRB versus the transmit power under different number of sensors. First, one can observe that the CRB of all the schemes monotonically decreases with $P_0$, since the received echo strength increases with the total transmit power, thereby improving the sensing performance. Second, it is observed that the CRB decreases by approximately 19 dB as the number of sensors increases from $L=2$ to $L=8$, which is consistent with the result in \eqref{sensing_crb}. This is due to the linear receive beamforming gain and the spatial direction gain of $\mathcal{O}(L^3)$ provided by the sensors. Third, it is observed that the CRB with the rotation angles optimized by the sensing-oriented scheme is close to that obtained by the ES scheme and outperforms those of the other schemes under different transmit power when $L=2$ and $L=8$. Notably, to achieve a CRB of -50 dB, the required transmit power is reduced from 24 dBm to 14 dBm by replacing the fixed rotation scheme with the proposed scheme when $L=8$, which demonstrates the effectiveness of the proposed design. This is due to the larger effective reception and reflection gains provided by rotation optimization, thereby maintaining the sensing performance at a lower transmit power.

\subsection{Single User Case}
\begin{figure}[t]
	\centering
	\includegraphics[width=0.85\linewidth]{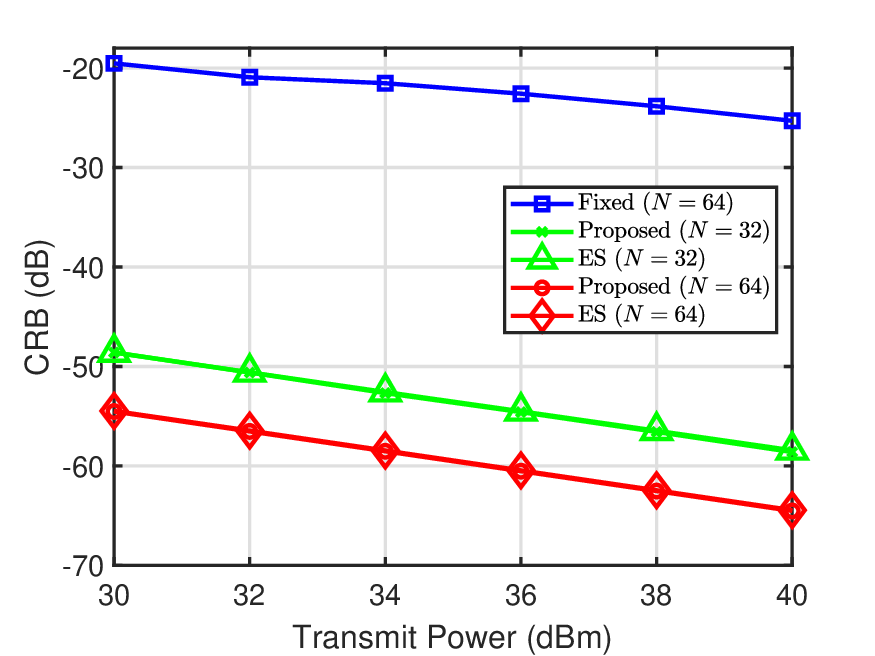}
	\vspace{-5pt}
	\caption{CRB versus the transmit power $P_0$.}
	\label{fig:STSU}
    \vspace{-5pt}
\end{figure}

Next, we consider the case where the BS serves one target and one user in the case with only a LoS-dominant path. In Fig. \ref{fig:STSU}, we plot the CRB versus the transmit power when $N=32$ and $N=64$. First, one can observe that the CRB of all schemes monotonically decreases with $P_0$, because increasing the transmit power budget improves sensing while satisfying the communication requirement. In particular, the CRB of the proposed scheme decreases by approximately 10 dB when $P_0$ increases from 30 dBm to 40 dBm for both $N=32$ and $N=64$. Second, it is observed that the CRB achieved by the proposed scheme is close to that of the ES scheme in the whole range of transmit power, which demonstrates the effectiveness of the proposed rotation design. In particular, both schemes outperform the fixed scheme even when the number of reflecting elements is halved ($N=32$ versus $N=64$). This is due to the effective aperture gains provided by rotation optimization. Furthermore, the CRB of the proposed scheme decreases by approximately 6 dB when the number of reflecting elements increases from $N=32$ to $N=64$. The results demonstrate that rotation optimization can provide additional DoFs and achieve better sensing performance with fewer reflecting elements, highlighting the importance of jointly designing the SIRS orientation and beamforming for ISAC.  

\begin{figure}[t]
	\centering
	\includegraphics[width=0.85\linewidth]{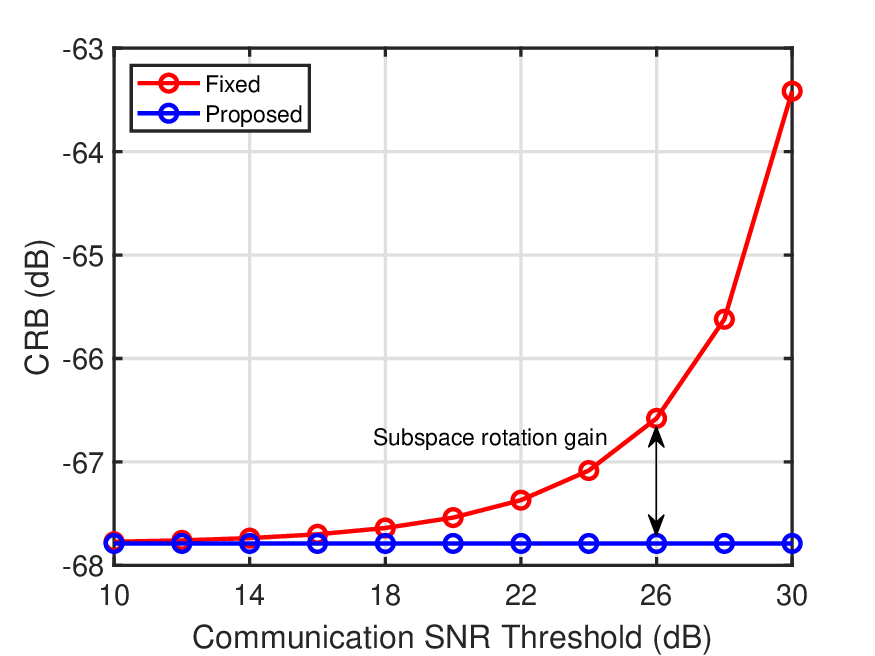}
	\vspace{-5pt}
	\caption{CRB versus the communication SNR threshold $\Gamma_{\mathrm{U}}$.}
	\label{fig:subspace}
    \vspace{-5pt}
\end{figure}

To further discuss the impact of the spatial correlation on the ISAC performance, we compare the proposed scheme based on Proposition \ref{proposition:subspace} and the fixed scheme by plotting the CRB versus the communication SNR threshold with $F_{\mathrm{T}}(\bm{\varphi}) \approx 1$ and $F_{\mathrm{U}}(\bm{\varphi}) \approx 1$ in Fig. \ref{fig:subspace}. First, one can observe that the CRB of the fixed scheme is close to that of the proposed scheme when the SNR threshold is low. This is because the SNR constraints are readily satisfied and thus most of the available spatial DoFs and transmit power can be allocated to sensing. However, the performance gap becomes larger as $\Gamma _\mathrm{U}$ increases, since the correlation between the effective sensing and communication channels is limited by the fixed orientation. Thus, the SIRS phase-shifts and rotation must be adjusted to improve the user signal strength at the expense of the beamforming gain towards the target, resulting in a higher CRB. In contrast, the CRB of the proposed scheme remains almost constant. Based on \eqref{lemma}, the rotation angles are optimized to increase $\rho$, thereby improving the effective sensing and communication channel gains through channel subspace rotation. The results demonstrate that the proposed design enlarges the region of the achievable  sensing and communication performance and facilitates high-quality sensing and communication services.

\subsection{Multi-User Case}
\begin{figure}[t]
	\centering
	\includegraphics[width=0.85\linewidth]{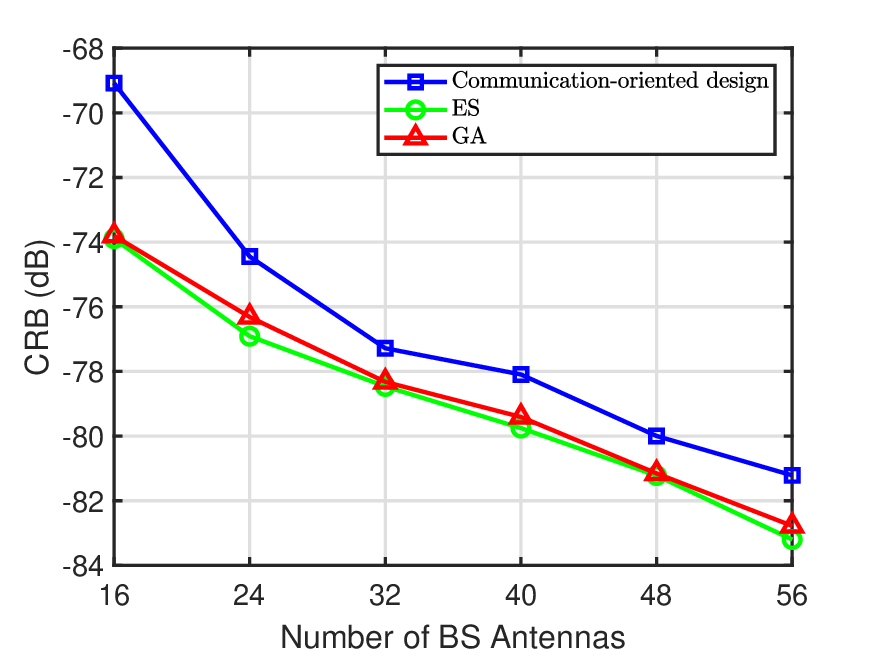}
	\vspace{-5pt}
	\caption{CRB versus the number of BS antennas $M$.}
	\label{fig:STMU_M}
    \vspace{-10pt}
\end{figure}

Finally, we study the case where the BS serves one target and two users in the ISAC system. Fig. \ref{fig:STMU_M} compares the CRB versus the number of BS antennas under the following SIRS rotation schemes: 1) Communication-oriented design: the SIRS rotation angles are set to maximize the minimum user SINR; 2) GA:  the optimized SIRS rotation angles are obtained via the genetic algorithm; 3) ES: the optimized SIRS rotation angles are obtained via the two-dimensional search. First, it is observed that the CRB of all the schemes monotonically decreases with the number of BS antennas due to the increased BS transmit beamforming gain. Second, we observe that the proposed GA scheme achieves a CRB close to that of the ES scheme in the whole range of $M$. Third, the CRB performance gap between the proposed scheme and the communication-oriented design scheme is pronounced when $M$ is relatively small. As $M$ increases, the transmit beamforming gain becomes dominant in achieving a balance between communication and sensing. Nevertheless, the gap remains noticeable at larger $M$. This is because the communication-oriented rotation scheme is aim to improve the communication performance to satisfy the SINR constraint, however, it does not account for the sensing information in the CRB. In contrast, the BS beamforming, the SIRS beamforming, and the SIRS rotation are jointly optimized under the GA and ES schemes to improve the ISAC performance. The results demonstrate the effectiveness of the GA scheme and highlight the importance of the SIRS rotation design and beamforming.

\begin{figure}[t]
	\centering
	\includegraphics[width=0.85\linewidth]{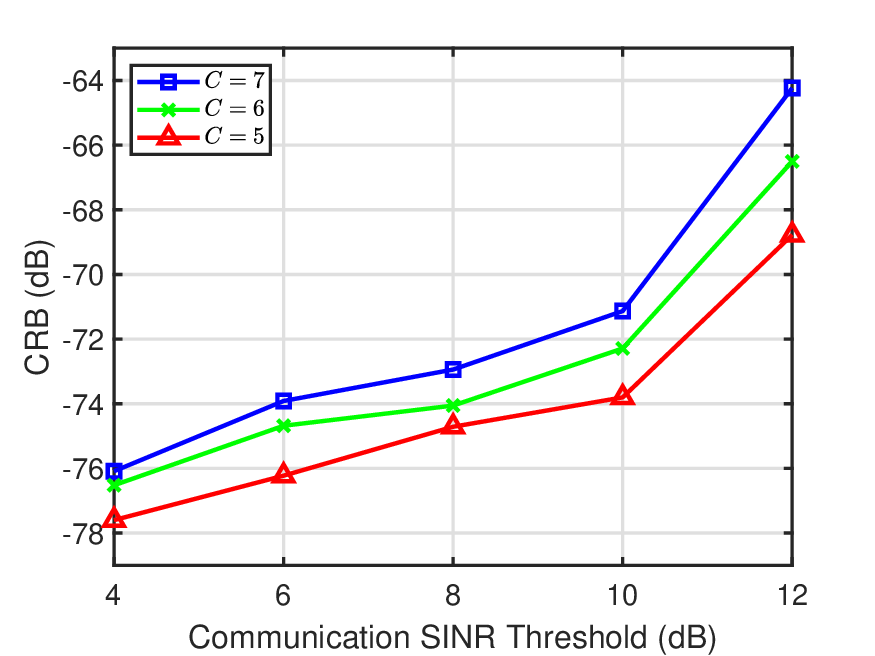}
	\vspace{-5pt}
	\caption{CRB versus the communication SINR threshold $\Gamma_{k}$.}
	\label{fig:STMU_C}
\end{figure}

To illustrate the fundamental trade-off between communication and sensing performance, we plot the CRB versus the communication SINR threshold in Fig. \ref{fig:STMU_C}. One can observe that the CRB increases with $\Gamma_k$. In particular, it increases more rapidly when the SINR threshold exceeds 10 dB. This is because the feasible region for the beamforming and rotation designs becomes more limited as the per-user communication SINR constraints become more stringent. In this case, more resources must be allocated to the communication signals to ensure sufficient received signal strength and suppress inter-user interference. Moreover, it is observed that increasing the number of scatterers leads to a higher CRB. Based on \eqref{path}, increasing $C$ reduces the weight of the LoS component, while the additional scattering paths experience cascaded propagation losses. However, the resulting path loss cannot be effectively compensated for by the additional spatial DoFs, especially when the SINR threshold is sufficiently large.

\begin{figure}[t]
	\centering
	\includegraphics[width=0.85\linewidth]{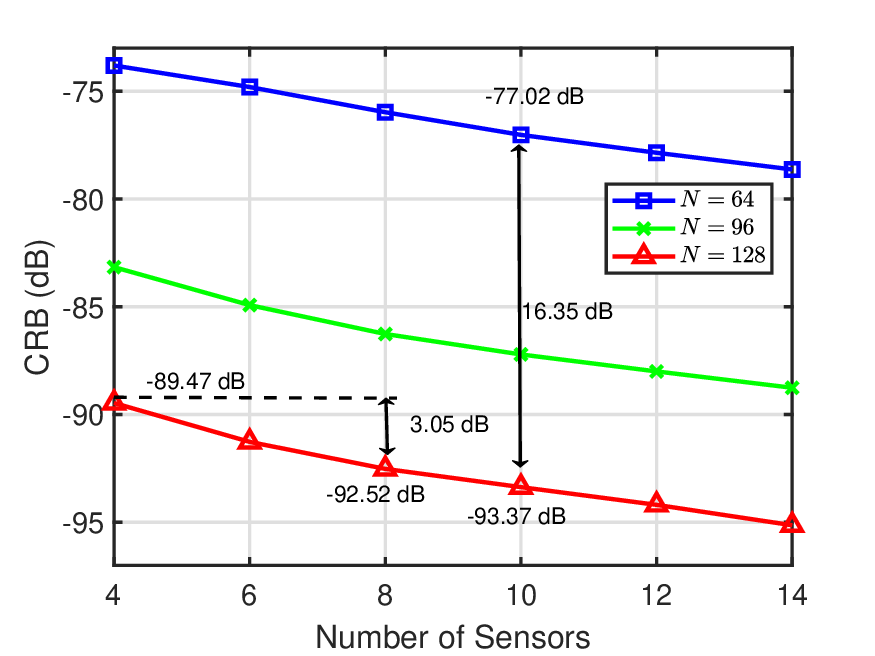}
	\vspace{-5pt}
	\caption{CRB versus the number of sensors $L$.}
	\label{fig:STMU_L}
\end{figure}

In the following, given the channel conditions, e.g., $C=5$, we investigate how to further improve the ISAC performance by exploiting the available system resources, including the numbers of SIRS reflecting elements and SIRS sensors. In Fig. \ref{fig:STMU_L}, we plot the CRB versus the number of sensors under different numbers of reflecting elements. We observe that the CRB monotonically decreases with $L$. This is because deploying more sensors at the SIRS provides higher spatial diversity gains, thereby increasing the information available for target parameter estimation. Different from the sensing-only case with ideal unit aperture gain, where the CRB is inversely proportional to $L^3-L$ and $N^2$ as shown in \eqref{sensing_crb}, increasing the number of reflecting elements is more beneficial for the considered system with multiple users. Specifically, the CRB decreases from -89.47 dB to -92.52 dB when $L$ increases from 4 to 8 with $N=128$, whereas the CRB decreases from -77.02 dB to -93.37 dB when $N$ increases from 64 to 128 with $L=10$. The reasons are as follows. In this case, the sensing performance depends not only on the improvement in target echo signal strength provided by the additional sensors but also on the communication constraints and the resulting joint beamforming and rotation designs. Moreover, additional reflecting elements provide a higher passive beamforming gain for both sensing and communication. With the joint optimization of beamforming and rotation, the spatial resources can be fully exploited to improve the sensing performance while satisfying the per-user communication SINR constraints. The results highlight the importance of determining the numbers of reflecting elements and sensors according to the specific sensing and communication constraints.

\section{Conclusion}
This paper has studied a rotatable SIRS-assisted ISAC system under an orientation-dependent model. Based on the proposed model, the sensing CRB was derived to characterize the effects of SIRS rotation and joint beamforming on target estimation. Closed-form expressions of the rotation angles were obtained for the sensing-only and communication-only cases, where two-dimensional orientation were sufficient to achieve same performance as three-dimensional orientation. For co-located user and target case, the sensing-oriented design satisfied the communication requirement. For the single-user single-target case with ideal unit aperture gains, the rotation angles were designed to improve the correlation between the effective communication and sensing channels. An effective method was developed to address the CRB minimization problem in the multi-user case. Numerical results demonstrated the effectiveness of the rotatable SIRS over the fixed-orientation SIRS in terms of reducing the requirement of transmit power and number of reflecting elements and improved the ISAC performance due to subspace rotation gain, and showed that it can improve the ISAC performance due to subspace rotation gain. Moreover, in the multi-user case, increasing the number of reflecting elements yields a larger CRB reduction than increasing the number of sensors. These findings provide useful guidelines for the joint design of surface orientation and beamforming in SIRS-aided ISAC systems.
	
\section*{Appendix A}
\section*{Proof of Proposition \ref{pro:2D}}
Based on \eqref{gCCS}, \eqref{theta_i}, and \eqref{vartheta_TI}, it is equivalent to prove
\begin{align}
	&\frac{{( {{x_{\mathrm{I}}} \!\!-\!\! {x_{\mathrm{T}}}} ) \! \cos {\varphi _x^{\star}} \! \sin {\varphi _y^{\star}} \!-\! ( {{y_{\mathrm{I}}} \!\!-\!\! {y_{\mathrm{T}}}} ) \! \sin {\varphi _x^{\star}} \!+\! ( {{z_{\mathrm{I}}} \!\!-\!\! {z_{\mathrm{T}}}} ) \! \cos {\varphi _x^{\star}} \! \cos {\varphi _y^{\star}}}}{{\sqrt {{{( {{x_{\mathrm{I}}} - {x_{\mathrm{T}}}} )}^2} + {{( {{y_{\mathrm{I}}} - {y_{\mathrm{T}}}} )}^2} + {{( {{z_{\mathrm{I}}} - {z_{\mathrm{T}}}} )}^2}} }} \nonumber\\
	&= \frac{{( {{x_{\mathrm{I}}} - {x_{\mathrm{T}}}} ){q_3} + ( {{y_{\mathrm{I}}} - {y_{\mathrm{T}}}} ){q_6} + ( {{z_{\mathrm{I}}} - {z_{\mathrm{T}}}} ){q_9}}}{{\sqrt {{{( {{x_{\mathrm{I}}} - {x_{\mathrm{T}}}} )}^2} + {{( {{y_{\mathrm{I}}} - {y_{\mathrm{T}}}} )}^2} + {{( {{z_{\mathrm{I}}} - {z_{\mathrm{T}}}} )}^2}} }}, \label{con:cosi}\\
	&\frac{{{x_{\mathrm{I}}}\cos {\varphi _x^{\star}}\sin {\varphi _y^{\star}} - {y_{\mathrm{I}}}\sin {\varphi _x^{\star}} + {z_{\mathrm{I}}}\cos {\varphi _x^{\star}}\cos {\varphi _y^{\star}}}}{{\sqrt {x_{\mathrm{I}}^2 + y_{\mathrm{I}}^2 + z_{\mathrm{I}}^2} }} \nonumber\\
	&= \frac{{{x_{\mathrm{I}}}{q_3} + {y_{\mathrm{I}}}{q_6} + {z_{\mathrm{I}}}{q_9}}}{{\sqrt {x_{\mathrm{I}}^2 + y_{\mathrm{I}}^2 + z_{\mathrm{I}}^2} }} \label{con:cost}.
\end{align}
The conditions under which \eqref{con:cosi} and \eqref{con:cost} hold are given by
\begin{align}
	&\cos {\varphi _x^{\star}}\sin {\varphi _y^{\star}} = q_3, \label{con1}\\
	&-\sin {\varphi _x^{\star}} = q_6, \label{con2}\\
	&\cos {\varphi _x^{\star}}\cos {\varphi _y^{\star}} = q_9. 
\end{align}

Next, we prove
\begin{align}
	& q_3^2+q_9^2 \le 1, \label{proof1}\\
	& |q_6| \le 1, \label{proof2}\\
	& q_3^2+q_6^2+q_9^2 = 1. \label{proof3}
\end{align}
Note that $\sin^2\varphi_x+\cos^2\varphi_x = 1$, it follows that
\begin{align}
	&1 \!-\! (q_3^2 \!+\! q_9^2) \! = \! \sin^2\varphi_x \!+\! \cos^2\varphi_x \!-\! \cos^2\varphi_x\sin^2\varphi_y\cos^2\varphi_z \nonumber\\
	&-\sin^2\varphi_x\sin^2\varphi_z-2\cos\varphi_x\sin\varphi_y\cos\varphi_z \nonumber\\
	&\times \sin\varphi_x\sin\varphi_z-\cos^2\varphi_x\cos^2\varphi_y \nonumber \\
	&=\sin^2\varphi_x\cos^2\varphi_z+\cos^2\varphi_x\sin^2\varphi_y\sin^2\varphi_z \nonumber\\
	&- 2\cos\varphi_x\sin\varphi_y\sin\varphi_z\sin\varphi_x\cos\varphi_z \nonumber\\
	&=(\cos\varphi_x\sin\varphi_y\sin\varphi_z-\sin\varphi_x\cos\varphi_z)^2 \ge 0.
\end{align}
The left hand side (LHS) of \eqref{proof2} can be rewritten as $q_6 = \left(\cos {\varphi _x}\sin {\varphi _y},-\sin {\varphi _x}\right) \left(\sin {\varphi _z},\cos {\varphi _z}\right)$.
Based on Cauchy–Schwarz inequality, we have 
\begin{align}
	|q_6| &\le \sqrt{\cos {\varphi _x}^2\sin {\varphi _y}^2+\sin {\varphi _x}^2}\sqrt{\sin {\varphi _z}^2+\cos {\varphi _z}^2} \nonumber\\
	&\le \sqrt{\cos {\varphi _x}^2\sin {\varphi _y}^2+\sin {\varphi _x}^2} \nonumber\\
	& \le \sqrt{\cos {\varphi _x}^2+\sin {\varphi _x}^2} = 1.
\end{align}
	
By substituting \eqref{gCCS} into the LHS of \eqref{proof3}, we have
\begin{align}
	q_3^2 + q_6^2 + q_9^2 = & \cos^2\varphi_x \sin^2\varphi_y (\cos^2\varphi_z + \sin^2\varphi_z) \nonumber\\
	&aa+ \sin^2\varphi_x (\sin^2\varphi_z + \cos^2\varphi_z) \nonumber\\
	= & \cos^2\varphi_x \sin^2\varphi_y + \sin^2\varphi_x + \cos^2\varphi_x \cos^2\varphi_y \nonumber\\
	= & \cos^2\varphi_x (\sin^2\varphi_y + \cos^2\varphi_y) + \sin^2\varphi_x \nonumber\\
	= & \cos^2\varphi_x + \sin^2\varphi_x = 1.
\end{align}
Based on \eqref{con1} and \eqref{con2}, we derive $\bm{\varphi}_\mathrm{2D}^{\star} = \left[\varphi_x^{\star}, \varphi_y^{\star}, 0\right]^T$ with
\begin{align}
	\varphi_x^{\star} = - \arcsin q_6, \;\; \varphi_y^{\star} = \arcsin \frac{q_3}{\sqrt {q_3^2 + q_9^2}}.
\end{align}
The proof is thus completed.

\section*{Appendix B}
\section*{Proof of Proposition \ref{pro:sensing}}
\vspace{-15pt}
\begin{align}
	\bar F \left(\varphi _x, \varphi _y\right) &= (\mathbf{q}_z^T \bar{\mathbf{p}}_{\mathrm{I}})(\mathbf{q}_z^T \bar{\mathbf{p}}_{\mathrm{T}}) \nonumber\\
	& = ((\mathbf{q}_z^T(\bar{\mathbf{p}}_{\mathrm{I}}+\bar{\mathbf{p}}_{\mathrm{T}}))^2 - (\mathbf{q}_z^T(\bar{\mathbf{p}}_{\mathrm{I}} - \bar{\mathbf{p}}_{\mathrm{T}}))^2) / 4 \nonumber\\
	&\mathop \le^{(a_1)} (\mathbf{q}_z^T(\bar{\mathbf{p}}_{\mathrm{I}}+\bar{\mathbf{p}}_{\mathrm{T}}))^2  / 4 \nonumber\\
	&\mathop \le^{(a_2)} \|\mathbf{q}_z^T\|^2 \|\bar{\mathbf{p}}_{\mathrm{I}}+ \bar{\mathbf{p}}_{\mathrm{T}}\|^2/4 \nonumber\\
	&= \|\bar{\mathbf{p}}_{\mathrm{I}}+ \bar{\mathbf{p}}_{\mathrm{T}}\|^2/4,
\end{align}
where $(a_1)$ and $(a_2)$ hold if and only if $\mathbf{q}_z^T \bar{\mathbf{p}}_{\mathrm{I}} = \mathbf{q}_z^T \bar{\mathbf{p}}_{\mathrm{T}}$ and $\mathbf{q}_z^T$ is parallel to $(\bar{\mathbf{p}}_{\mathrm{I}}+ \bar{\mathbf{p}}_{\mathrm{T}})$, respectively. Thus, when $\bar F \left(\bm{\varphi}\right)$ is maximized, we have 
\begin{align}
	\mathbf{q}_z^\star = \frac{\bar{\mathbf{p}}_{\mathrm{I}}+ \bar{\mathbf{p}}_{\mathrm{T}}}{\|\bar{\mathbf{p}}_{\mathrm{I}}+ \bar{\mathbf{p}}_{\mathrm{T}}\|} = \frac{\mathbf{\Delta}_1}{\|\mathbf{\Delta}_1\|},
\end{align}
Then, we have
\begin{align}
	\label{qtophi}
	&\cos \varphi_x^\star \sin \varphi_y^\star = \frac{\mathbf{\Delta}_{1,x}}{\|\mathbf{\Delta}_{1}\|},\nonumber\\
	&- \sin \varphi_x^\star = \frac{\mathbf{\Delta}_{1,y}}{\|\mathbf{\Delta}_{1}\|},\nonumber\\
	&\cos \varphi_x^\star \cos \varphi_y^\star = \frac{\mathbf{\Delta}_{1,z}}{\|\mathbf{\Delta}_{1}\|},
\end{align}
As such, the optimal rotation angles are given by
\begin{align}
	\varphi_x^\star = -\arcsin \frac{\mathbf{\Delta}_{1,y}}{\|\mathbf{\Delta}_{1}\|}, \;\;
	\varphi_y^\star = \arctan \frac{\mathbf{\Delta}_{1,x}}{\mathbf{\Delta}_{1,z}},
\end{align}
which completes the proof.

\section*{Appendix C}
\section*{Proof of Proposition \ref{proposition:co}}

Let $b_1 = {\cos {\theta _{\mathrm{I}}}}$ and $b_2 = {{\cos }} {\theta _{\mathrm{T}}^{\mathrm{r}}}$. Introducing the auxiliary variables $\bar \beta$, which denotes the angle between $\mathbf{q}_z$ and $\mathbf{\bar{p}}_{\mathrm T}$, we have $b_1 = \cos(\bar \alpha - \bar  \beta)$ and $b_2 = \cos \bar \beta$. By leveraging the addition formula for Cosine, it follows that $b_1 = \cos \bar \alpha \cos \bar \beta + \sin \bar \alpha \sin \bar \beta = b_2 \cos \bar \alpha + \sin \bar \alpha \sqrt{1 - b_2^2}$. Then, the resulting equality constraint used in the Lagrangian function is given by $\sin^2 \bar \alpha - b_1^2 - b_2^2 + 2b_1b_2 \cos \bar \alpha = 0$. As such, problem \eqref{pro:co} can be reformulated as
\begin{subequations}
	\label{pro:co1}
	\begin{align}
		\mathop {\max }\limits_{b_1, b_2} \;\;\; &b_1 b_2 \\
		\mathrm{s.t.} \;\;\;\;
		& b_1 b_2 {|\alpha _{{\mathrm{U}}}|^2}{|\alpha _{{\mathrm{BI}}}|^2}P_0N^2M/{\sigma _{{\mathrm{U}}}^2} \ge \Gamma, \\
		& \sin^2 \bar \alpha - b_1^2 - b_2^2 + 2b_1b_2 \cos \bar \alpha = 0, \label{equ_b1b2} \\
		& 0 \le b_1 \le 1, 0 \le b_2 \le 1.
	\end{align}
\end{subequations}
Based on \eqref{equ_b1b2}, we have
\begin{align}
	b_1 b_2 &= \frac{1 + \cos \bar{\alpha}}{2} - \frac{(b_1 - b_2)^2}{2(1 - \cos \bar{\alpha})} \mathop \le^{(a_3)} \frac{1 + \cos \bar{\alpha}}{2},
\end{align}
where $(a_3)$ holds if and only if $b_1 = b_2$. Then, it follows that $b_1^\star = b_2^\star = \sqrt{\frac{1 + \cos \bar{\alpha}}{2}}$ if $\cos \bar{\alpha} \ge 2 \bar \Gamma - 1$. With $b_1$ and $b_2$, we have
\begin{align}
	\mathbf{q}_z^\star &= \frac{b_1^\star - b_2^\star \cos \bar{\alpha}}{\sin^2 \bar{\alpha}} \bar{\mathbf{p}}_{\mathrm{I}} + \frac{b_2^\star - b_1^\star \cos \bar{\alpha}}{\sin^2 \bar{\alpha}} \bar{\mathbf{p}}_{\mathrm{T}} \nonumber\\
	&= \frac{\bar{\mathbf{p}}_{\mathrm{I}} + \bar{\mathbf{p}}_{\mathrm{T}}}{\sqrt{2(1 + \cos \bar{\alpha})}} = \frac{\mathbf{\Delta}_1}{\|\mathbf{\Delta}_1\|}.
\end{align}
As such, the optimal rotation angles are given by
\begin{align}
	\varphi_x^\star = -\arcsin \frac{\mathbf{\Delta}_{1,y}}{\|\mathbf{\Delta}_1\|}, 
	\varphi_y^\star = \arctan \frac{\mathbf{\Delta}_{1,x}}{\mathbf{\Delta}_{1,z}},
\end{align}
which completes the proof.

\section*{Appendix D}
\section*{Proof of Proposition \ref{proposition:subspace}}

With a standard Dirichlet function, we can rewrite $g_1 = \rho \sqrt{\frac{\bar{\Gamma}_{\mathrm{U}}}{\|\mathbf{u}_{\mathrm{U}}\|^2}} + \sqrt{N^2 - \frac{\bar{\Gamma}_{\mathrm{U}}}{\|\mathbf{u}_{\mathrm{U}}\|^2}} \sqrt{1 - \rho^2}$ into a compact form as 
\begin{align}
	\rho  = \left| \frac{\sin\left( N \frac{\pi d_{\mathrm{I}}}{\lambda} (\sin\theta_{\mathrm{I,U}} - \sin\vartheta_{\mathrm{TI}}) \right)}{N\sin\left( \frac{\pi d_{\mathrm{I}}}{\lambda} (\sin\theta_{\mathrm{I,U}} - \sin\vartheta_{\mathrm{TI}}) \right)} \right|.
\end{align}
Hence, $\sin\theta_{\mathrm{I,U}} = \sin\vartheta_{\mathrm{TI}}$ is sufficient for $\rho = 1$, which yields $\mathbf{q}_z^T \mathbf{c}_\mathrm{UT} = 0$. To satisfy the constraints, the optimized normal vector $\mathbf{q}_z^* = \frac{\mathbf{\Delta}_{3}}{\|\mathbf{\Delta}_3\|}$ belongs to the feasible set $\mathcal{Q} = \{ \mathbf{q}_z | \mathbf{q}_z^T \mathbf{c}_\mathrm{UT} \!=\! 0, \|\mathbf{q}_z\| \!=\! 1, \mathbf{q}_z^T \bar{\mathbf{p}}_\mathrm{I} \! \ge \! 0, \mathbf{q}^T \bar{\mathbf{p}}_\mathrm{T} \! \ge \! 0, \mathbf{q}_z^T \bar{\mathbf{p}}_\mathrm{U} \! \ge \! 0 \}$. 


The optimized SIRS phase shift can be expressed as $\phi _n^* ={\arg {{[{\bf{\tilde v}}_{{\mathrm{opt}}}^*]}_n}}$,
where $\mathbf{\tilde{v}}_{\mathrm{opt}} = x_1 \mathbf{e}_1 + x_2 \mathbf{e}_2$ with $x_1 = \sqrt{\frac{\bar{\Gamma}_{\mathrm{U}}}{ \|\mathbf{u}_{\mathrm{U}}\|^2}} \frac{\mathbf{e}_1^H \mathbf{u}_{\mathrm{T}}}{|\mathbf{e}_1^H \mathbf{u}_{\mathrm{T}}|}$, $x_2 = \sqrt{N - \frac{\bar{\Gamma}_{\mathrm{U}}}{ \|\mathbf{u}_{\mathrm{U}}\|^2}} \frac{\mathbf{e}_2^H \mathbf{u}_{\mathrm{T}}}{|\mathbf{e}_2^H \mathbf{u}_{\mathrm{T}}|}$, $\mathbf{e}_1 = \frac{\mathbf{u}_{\mathrm{U}}}{\|\mathbf{u}_{\mathrm{U}}\|}$, $\mathbf{e}_2 = \frac{\mathbf{u}_{\mathrm{T}} - (\mathbf{e}_1^H \mathbf{u}_{\mathrm{T}})\mathbf{e}_1}{\|\mathbf{u}_{\mathrm{T}} - (\mathbf{e}_1^H \mathbf{u}_{\mathrm{T}})\mathbf{e}_1\|}$. Since $\rho  \le  1$ and $g_1$ in \eqref{lemma} is upper bounded by $N^2$, the optimal SIRS phase shift is given by $\phi _n^* = -{\arg {{\left[ {{{\bf{a}}_{{\mathrm{IT}}}}\left( {{\vartheta _{{\mathrm{TI}}}}} \right)} \right]}_n} \!-\! \arg {{\left[ {{{\bf{a}}_{{\mathrm{IB}}}}\left( {{\vartheta _{{\mathrm{IB}}}}} \right)} \right]}_n}}$,
which completes the proof.

\bibliographystyle{IEEEtran}
\bibliography{refs.bib} 

@INPROCEEDINGS{rotation_IRS,
	author={Xu, De and Wang, Tao and Yang, Nuocheng and He, Xinxin and Yin, Changchuan and You, Changsheng and Tong, Haonan},
	title={{IRS} Deployment Under Practical Channel Model: Joint Position and Rotation Optimization},
	booktitle={Proc. 2025 11th Int. Conf. Comput. Commun. (ICCC)},
	pages={981--986},
	year={2025},
	month={Dec.},
	doi={10.1109/ICCC68654.2025.11437992}
}

@ARTICLE{songISAC,
	author={Song, Xianxin and Qin, Xiaoqi and Xu, Jie and Zhang, Rui},
	title={Cram{\'e}r-Rao Bound Minimization for {IRS}-Enabled Multiuser Integrated Sensing and Communications},
	journal={IEEE Trans. Wireless Commun.},
	volume={23},
	number={8},
	pages={9714--9729},
	year={2024},
	month={Aug.},
	doi={10.1109/TWC.2024.3364913}
}

@INPROCEEDINGS{S&C,
	author={Meng, Xiao and Liu, Fan and Lu, Shihang and Chepuri, Sundeep Prabhakar and Masouros, Christos},
	title={{RIS}-assisted Integrated Sensing and Communications: A Subspace Rotation Approach: Invited Paper},
	booktitle={Proc. 2023 IEEE Radar Conference (RadarConf23)},
	pages={1--6},
	year={2023},
	month={May},
	doi={10.1109/RadarConf2351548.2023.10149664}
}

@ARTICLE{beam,
	author={Huang, Yongwei and Palomar, Daniel P.},
	title={Rank-Constrained Separable Semidefinite Programming With Applications to Optimal Beamforming},
	journal={IEEE Trans. Signal Process.},
	volume={58},
	number={2},
	pages={664--678},
	year={2010},
	month={Feb.},
	doi={10.1109/TSP.2009.2031732}
}

@ARTICLE{RA_UAV,
	author={Chen, Shiying and Chen, Guangji and Shi, Long and Wu, Qingqing and Wei, Kang},
	title={Rotatable Antenna Meets {UAV}: Toward Dual-Level Channel Reconfiguration Paradigm for {ISAC}},
	journal={IEEE Trans. Veh. Technol.},
	volume={75},
	number={9},
	pages={21760--21765},
	year={2026},
	month={Sep.},
	doi={10.1109/TVT.2026.3682292}
}

@ARTICLE{luo2026ISACSurvey,
	author={Luo, Xuewen and Lin, Qingfeng and Zhang, Ruoyu and Chen, Hsiao-Hwa and Wang, Xingwei and Huang, Min},
	title={{ISAC}---A Survey on Its Layered Architecture, Technologies, Standardizations, Prototypes, and Testbeds},
	journal={IEEE Commun. Surveys Tuts.},
	volume={28},
	pages={485--526},
	year={2026},
	doi={10.1109/COMST.2025.3565534}
}

@ARTICLE{hua2024MIMOTradeoff,
	author={Hua, Haocheng and Han, Tony Xiao and Xu, Jie},
	title={{MIMO} Integrated Sensing and Communication: {CRB}-Rate Tradeoff},
	journal={IEEE Trans. Wireless Commun.},
	volume={23},
	number={4},
	pages={2839--2854},
	year={2024},
	month={Apr.},
	doi={10.1109/TWC.2023.3303326}
}

@ARTICLE{ren2024MultitargetTradeoff,
	author={Ren, Zixiang and Peng, Yunfei and Song, Xianxin and Fang, Yuan and Qiu, Ling and Liu, Liang and Ng, Derrick Wing Kwan and Xu, Jie},
	title={Fundamental {CRB}-Rate Tradeoff in Multi-Antenna {ISAC} Systems With Information Multicasting and Multi-Target Sensing},
	journal={IEEE Trans. Wireless Commun.},
	volume={23},
	number={4},
	pages={3870--3885},
	year={2024},
	month={Apr.},
	doi={10.1109/TWC.2023.3312723}
}

@ARTICLE{bian2025Correlation,
	author={Bian, Mengqi and Shi, Yunmei and Huang, Xin-Lin},
	title={Fundamental {CRB}-Rate Trade-Off in {ISAC} Systems Under Correlated Communication-Sensing Channel},
	journal={IEEE Trans. Commun.},
	volume={73},
	number={11},
	pages={10724--10737},
	year={2025},
	month={Nov.},
	doi={10.1109/TCOMM.2025.3588613}
}

@ARTICLE{tishchenko2025HybridSurvey,
	author={Tishchenko, Anton and others},
	title={The Emergence of Multi-Functional and Hybrid Reconfigurable Intelligent Surfaces for Integrated Sensing and Communications - A Survey},
	journal={IEEE Commun. Surveys Tuts.},
	volume={27},
	number={5},
	pages={2895--2936},
	year={2025},
	month={Oct.},
	doi={10.1109/COMST.2024.3519785}
}

@ARTICLE{niu2025InterferenceSurvey,
	author={Niu, Yangyang and Wei, Zhiqing and Wang, Lin and Wu, Huici and Feng, Zhiyong},
	title={Interference Management for Integrated Sensing and Communication Systems: A Survey},
	journal={IEEE Internet Things J.},
	volume={12},
	number={7},
	pages={8110--8134},
	year={2025},
	month={Apr.},
	doi={10.1109/JIOT.2024.3506162}
}

@ARTICLE{rihan2024PassiveActive,
	author={Rihan, Mohamed and Zappone, Alessio and Buzzi, Stefano and Fodor, Gabor and Debbah, M{\'e}rouane},
	title={Passive Versus Active Reconfigurable Intelligent Surfaces for Integrated Sensing and Communication: Challenges and Opportunities},
	journal={IEEE Netw.},
	volume={38},
	number={3},
	pages={218--226},
	year={2024},
	month={May},
	doi={10.1109/MNET.2023.3321542}
}

@ARTICLE{sankar2024RISBeamforming,
	author={Sankar, R. S. Prasobh and Chepuri, Sundeep Prabhakar and Eldar, Yonina C.},
	title={Beamforming in Integrated Sensing and Communication Systems With Reconfigurable Intelligent Surfaces},
	journal={IEEE Trans. Wireless Commun.},
	volume={23},
	number={5},
	pages={4017--4031},
	year={2024},
	month={May},
	doi={10.1109/TWC.2023.3313938}
}

@ARTICLE{liu2024SNRCRB,
	author={Liu, Rang and Li, Ming and Liu, Qian and Lee Swindlehurst, A.},
	title={{SNR}/{CRB}-Constrained Joint Beamforming and Reflection Designs for {RIS-ISAC} Systems},
	journal={IEEE Trans. Wireless Commun.},
	volume={23},
	number={7},
	pages={7456--7470},
	year={2024},
	month={Jul.},
	doi={10.1109/TWC.2023.3341429}
}

@ARTICLE{zhu2024ActiveCRB,
	author={Zhu, Qi and Li, Ming and Liu, Rang and Liu, Qian},
	title={Cram{\'e}r-Rao Bound Optimization for Active {RIS}-Empowered {ISAC} Systems},
	journal={IEEE Trans. Wireless Commun.},
	volume={23},
	number={9},
	pages={11723--11736},
	year={2024},
	month={Sep.},
	doi={10.1109/TWC.2024.3384501}
}

@ARTICLE{peng2024SemiPassive,
	author={Peng, Qiaoyan and Wu, Qingqing and Chen, Wen and Ma, Shaodan and Zhao, Ming-Min and Dobre, Octavia A.},
	title={Semi-Passive Intelligent Reflecting Surface-Enabled Sensing Systems},
	journal={IEEE Trans. Commun.},
	volume={72},
	number={12},
	pages={7674--7688},
	year={2024},
	month={Dec.},
	doi={10.1109/TCOMM.2024.3399202}
}

@ARTICLE{dong2025SemiPassiveNearField,
	author={Dong, Xiequn and Fei, Zesong and Hua, Meng and Wang, Xinyi and Liu, Yuanwei},
	title={Semi-Passive {RIS} Aided Near-Field {ISAC}: {CRB} Analysis and Optimization},
	journal={IEEE Trans. Veh. Technol.},
	volume={74},
	number={6},
	pages={9084--9099},
	year={2025},
	month={Jun.},
	doi={10.1109/TVT.2025.3534287}
}

@ARTICLE{rojith2026SemiPassiveEE,
	author={K, Rojith and Yu, Chih-Yuan and Singh, Keshav and Li, Chih-Peng},
	title={Joint Beamforming and Reflection Design Toward Energy Efficiency Maximization in Semi-Passive {RIS} Assisted {ISAC} Systems},
	journal={IEEE Trans. Veh. Technol.},
	volume={75},
	number={10},
	pages={24010--24023},
	year={2026},
	month={Oct.},
}

@ARTICLE{hua2024SemiPassiveLocalization,
	author={Hua, Meng and others},
	title={Intelligent Reflecting Surface-Assisted Localization: Performance Analysis and Algorithm Design},
	journal={IEEE Wireless Commun. Lett.},
	volume={13},
	number={1},
	pages={84--88},
	year={2024},
	month={Jan.},
	doi={10.1109/LWC.2023.3320728}
}

@ARTICLE{dong2026AngleVelocity,
	author={Dong, Xiequn and Fei, Zesong and Hua, Meng and Huang, Jingxuan and Wang, Xinyi and Xiao, Yue},
	title={Reconfigurable Intelligent Surfaces for {ISAC}: {CRB} Analysis and Optimization for Joint Angle and Radial Velocity Estimation},
	journal={IEEE Trans. Veh. Technol.},
	volume={75},
	number={2},
	pages={2396--2410},
	year={2026},
	month={Feb.},
	doi={10.1109/TVT.2025.3598365}
}

@ARTICLE{liu2025RotatableIRS,
	author={Liu, Jianghui and Zhang, Hongtao},
	title={Max-Min Area Coverage in {IRS} Assisted Networks: Rotatable Design and Experimental Verification},
	journal={IEEE Trans. Commun.},
	volume={73},
	number={11},
	pages={12355--12365},
	year={2025},
	month={Nov.},
	doi={10.1109/TCOMM.2025.3571960}
}

@ARTICLE{feng2025SixDIRS,
	author={Feng, Jie and Zheng, Beixiong and You, Changsheng and Chen, Peng and Xiong, Xue and Chen, Fangjiong},
	title={{6-D} Deployment Optimization for Intelligent Reflecting Surface-Aided Wireless Communication},
	journal={IEEE Wireless Commun. Lett.},
	volume={14},
	number={5},
	pages={1276--1280},
	year={2025},
	month={May},
	doi={10.1109/LWC.2024.3524592}
}

@ARTICLE{qu2025RotatableCooperativeISAC,
	author={Qu, Kaiqian and Li, Haojin and Sun, Chen and Zhang, Wenqi and Guo, Shuaishuai and Zhang, Haijun},
	title={Rotatable Array-Enabled Multi-{BS} Cooperative {ISAC} Transmit Beampattern Design},
	journal={IEEE Trans. Veh. Technol.},
	volume={74},
	number={9},
	pages={14775--14780},
	year={2025},
	month={Sep.},
	doi={10.1109/TVT.2025.3559665}
}

@ARTICLE{wang2026RotatableHybridISAC,
	author={Wang, Zequan and Yin, Liang and Lei, Zimeng and Liu, Yitong and Sun, Yunan and Yang, Hongwen},
	title={Rotatable Array-Aided Hybrid Beamforming for Integrated Sensing and Communication},
	journal={IEEE Internet Things J.},
	volume={13},
	number={7},
	pages={14364--14377},
	year={2026},
	month={Apr.},
	doi={10.1109/JIOT.2025.3649203}
}

@ARTICLE{zhang2026RotatableNearField,
	author={Zhang, Yunpu and others},
	title={Rotatable Antennas for Near-Field Integrated Sensing and Communication},
	journal={IEEE Trans. Wireless Commun.},
	volume={25},
	pages={10986--11001},
	year={2026},
	doi={10.1109/TWC.2026.3650918}
}

@ARTICLE{sun2025RotatableMovableISAC,
	author={Sun, Yunan and Xu, Hao and Ouyang, Chongjun and Yang, Hongwen},
	title={Rotatable and Movable Antenna-Enabled Near-Field Integrated Sensing and Communication},
	journal={IEEE Internet Things J.},
	volume={12},
	number={21},
	pages={45119--45132},
	year={2025},
	month={Nov.},
	doi={10.1109/JIOT.2025.3598721}
}

@ARTICLE{yigit2025HybridSTARISAC,
	author={Yigit, Zehra and Basar, Ertugrul},
	title={Hybrid {STAR-RIS} Enabled Integrated Sensing and Communication},
	journal={IEEE Trans. Commun.},
	volume={73},
	number={9},
	pages={8289--8300},
	year={2025},
	month={Sep.},
	doi={10.1109/TCOMM.2025.3547800}
}

@ARTICLE{han2025MultifunctionalRIS,
	author={Han, Dongsheng and others},
	title={Multi-Functional {RIS} Integrated Sensing and Communications for {6G} Networks},
	journal={IEEE Trans. Wireless Commun.},
	volume={24},
	number={2},
	pages={1146--1161},
	year={2025},
	month={Feb.},
	doi={10.1109/TWC.2024.3505918}
}

@ARTICLE{hua2024SecureIRSISAC,
	author={Hua, Meng and others},
	title={Secure Intelligent Reflecting Surface-Aided Integrated Sensing and Communication},
	journal={IEEE Trans. Wireless Commun.},
	volume={23},
	number={1},
	pages={575--591},
	year={2024},
	month={Jan.},
	doi={10.1109/TWC.2023.3280179}
}

@ARTICLE{hu2023LocationISAC,
	author={Hu, Xiaoling and Liu, Chenxi and Peng, Mugen and Zhong, Caijun},
	title={{IRS}-Based Integrated Location Sensing and Communication for {mmWave} {SIMO} Systems},
	journal={IEEE Trans. Wireless Commun.},
	volume={22},
	number={6},
	pages={4132--4145},
	year={2023},
	month={Jun.},
	doi={10.1109/TWC.2022.3223428}
}

@ARTICLE{luo2026MultihopPositioning,
	author={Luo, Yirui and Ma, Xiaoyan and Guan, Yong Liang and Brinton, Christopher G. and Yuen, Chau},
	title={Multi-Hop {RIS} {ISAC} for Target Positioning: {A} Tensor Decomposition-Based Approach},
	journal={IEEE Trans. Wireless Commun.},
	volume={25},
	pages={21545--21561},
	year={2026},
	doi={10.1109/TWC.2026.3721246}
}

@ARTICLE{ma2024RISISAC,
	author={Ma, Teng and Xiao, Yue and Lei, Xia and Renzo, Marco Di},
	title={Integrated Sensing and Communication With Reconfigurable Intelligent Surfaces},
	journal={IEEE Trans. Veh. Technol.},
	volume={73},
	number={12},
	pages={19051--19064},
	year={2024},
	month={Dec.},
	doi={10.1109/TVT.2024.3442574}
}

@ARTICLE{peng2026CooperativeRotatableIRS,
	author={Peng, Qiaoyan and Wu, Qingqing and Chen, Guangji and Chen, Wen and Shen, Shanpu and Ma, Shaodan},
	title={Cooperative Rotatable {IRSs} for Wireless Communications: Joint Beamforming and Orientation Optimization},
	journal={IEEE Trans. Veh. Technol.},
	volume={75},
	number={10},
	pages={25343--25348},
	year={2026},
	month={Oct.},
	doi={10.1109/TVT.2026.3701223}
}

@ARTICLE{peng2026ActiveULDL,
	author={Peng, Qiaoyan and Wu, Qingqing and Chen, Guangji and Chen, Wen and Ma, Shaodan},
	title={Active {IRS}-Assisted Joint Uplink and Downlink Communications},
	journal={IEEE Trans. Wireless Commun.},
	volume={25},
	pages={19873--19887},
	year={2026},
	doi={10.1109/TWC.2026.3711180}
}

@article{selfSensingIRS,
  author       = {Zijian Chen and others},
  title        = {Joint Location Sensing and Channel Estimation for {IRS}-Aided mmWave
                  {ISAC} Systems},
  journal      = {{IEEE} Trans. Wireless Commun.},
  volume       = {23},
  number       = {9},
  pages        = {11985--12002},
  year         = {2024},
  month = {Dec.}
}

@article{wang2025rotate,
  author       = {Peilan Wang and
                  Yu Xue and
                  Weidong Mei and
                  Jun Fang and
                  Rui Zhang},
  title        = {UAV-Enabled Passive {6D} Movable Antenna for {ISAC:} Joint Location,
                  Orientation, and Reflection Optimization},
  journal      = {{IEEE} Wireless Commun. Lett.},
  volume       = {14},
  number       = {12},
  pages        = {3982--3986},
  year         = {2025},
  month        = {Sep.}
}

@article{chen2026ISAC,
  author       = {Guangji Chen and
                  Qingqing Wu and
                  Shihang Lu and
                  Meng Hua and
                  Wen Chen},
  title        = {Multi-{IRS}-Aided {ISAC} System: Multi-Path Exploitation Versus Reduction},
  journal      = {{IEEE} Trans. Wireless Commun.},
  volume       = {25},
  pages        = {13351--13368},
  year         = {2026},
  month        = {Mar.}
}

@article{feng2026ISAC,
  author       = {Haotian Liu and
                  Zhiqing Wei and
                  Luyang Sun and
                  Ruizhong Xu and
                  Yixin Zhang and
                  Zhiyong Feng},
  title        = {Cooperative Sensing in Cell-Free Massive {MIMO} {ISAC} Systems: Performance
                  Optimization and Signal Processing},
  journal      = {{IEEE} Trans. Wireess Commun.},
  volume       = {25},
  pages        = {12531--12547},
  year         = {2026},
  month        = {Jan.}
}

@article{luRA,
  author       = {Haiquan Lu and
                  Zhi Yu and
                  Yong Zeng and
                  Shaodan Ma and
                  Shi Jin and
                  Rui Zhang},
  title        = {Wireless Communication With Flexible Reflector: Joint Placement and Rotation Optimization for Coverage Enhancement},
  journal      = {{IEEE} Trans. Wireless Commun.},
  volume       = {24},
  number       = {10},
  pages        = {8252--8266},
  year         = {2025},
  month={Oct.}
}

@article{huangUAV,
  author       = {Jiapeng Li and
                  Qixun Zhang and
                  Dingyou Ma and
                  Sai Huang},
  title        = {{UAV} Target Reconstruction and Imaging Algorithm Design and Performance
                  Evaluation Using {CSI} of {ISAC} Signal},
  journal      = {{IEEE} Trans. Wireless Commun.},
  volume       = {24},
  number       = {11},
  pages        = {9422--9434},
  year         = {2025},
  month={Nov.}
}

@article{colocated,
  author       = {Shihang Lu and
                  Fan Liu and
                  Lajos Hanzo},
  title        = {The Degrees-of-Freedom in Monostatic {ISAC} Channels: {NLoS} Exploitation vs. Reduction},
  journal      = {{IEEE} Trans. Veh. Technol.},
  volume       = {72},
  number       = {2},
  pages        = {2643--2648},
  year         = {2023},
  month={Jan.}
}

@ARTICLE{ji_RDARS_model_driven,
  author={Ji, Chengwang and others},
  journal={IEEE Trans. Cogn. Commun. Netw.}, 
  title={Model-Driven Deep Learning Enhanced Joint Beamforming and Mode Switching for {RDARS}-Aided {MIMO} Systems}, 
  year={2026},
  volume={12},
  pages={13109-13124},
  month = {Oct.},
}
	
\end{document}